\documentclass[fleqn,usenatbib]{mnras}
\usepackage{newtxtext,newtxmath}
\usepackage[T1]{fontenc}
\DeclareRobustCommand{\VAN}[3]{#2}
\let\VANthebibliography\thebibliography
\def\thebibliography{\DeclareRobustCommand{\VAN}[3]{##3}\VANthebibliography}

\usepackage{graphicx}	
\usepackage{amsmath}	
\usepackage{multirow}
\usepackage{longtable}
\usepackage{graphicx}

\usepackage{xcolor}

\title[Transient quasi-periodic oscillations in the gamma-ray light curves]{Transient quasi-periodic oscillations in the gamma-ray light curves of bright blazars}

\author[J. P. Chen et al.]{
Junping Chen$^{1,2}$,
Jinjie Yu$^{1}$,
Weitian Huang$^{3}$,
and Nan Ding$^{1}$\thanks{Corresponding author: orient.dn@foxmail.com}
\\
$^{1}$School of Physical Science and Technology, Kunming University, Kunming, 650214, People’s Republic of China\\
$^{2}$School of Physics and Astronomy, Sun Yat-Sen University, Zhuhai, 519000, People's Republic of China\\
$^{3}$Department of Physics, Yunnan Normal University, Kunming, 650092, People's Republic of China\\
}
\date{Accepted XXX. Received YYY; in original form ZZZ}
\pubyear{2023}
\begin{document}
\label{firstpage}
\pagerange{\pageref{firstpage}--\pageref{lastpage}}
\maketitle

\begin{abstract}
Transient quasi-periodic oscillations (QPOs) are extremely interesting observational phenomena. However, the precise physical mechanisms leading to their generation are still hotly debated. We performed a systematic search for transient QPO signals using Weighted Wavelet Z-transforms on the gamma-ray light curves of 134 bright blazars with peak flux exceeding $1\times10^{-6}$~ph~cm$^{-2}$~s$^{-1}$ as monitored by Fermi-LAT. Artificial light curves were generated from the power spectral density and probability distribution functions of the original light curves to assess the significance level of transient QPO. We discuss several physical mechanisms that produce transient QPOs, with the helical jet model providing the best explanation. This study identified four new transient QPO events. Interestingly, repetitive transient QPOs are observed in PKS 0537-441, and nested transient QPOs are detected in PKS 1424-41. 
Additionally, we find that transient QPOs tend to occur in the flare state of the blazar. Finally, we estimate the incidence of transient QPO events to be only about 3\%. 
\end{abstract}


\section{Introduction}
There are usually one or more supermassive black holes in the center of active galactic nuclei (the mass is millions to billions of times the mass of the sun), and the supermassive black hole accretes surrounding matter into bright radiation. There are many subclasses of active galactic nuclei, the type in which the jet is directed toward the observer is called a blazar \citep{1995PASP..107..803U,2019ARA&A..57..467B}.
Blazars are also among the brightest objects in the universe, with radiation spanning the entire electromagnetic spectrum and significant long-term light variability. The energy spectrum of blazars shows a double-peak structure in the electromagnetic band, the low-energy peak is dominated by optical to X-ray emission, and the high-energy peak is mainly dominated by gamma-ray emission. Current studies have shown that low-energy photons mainly come from synchrotron radiation, while high-energy photons mainly come from inverse Compton scattering, also originated from the relativistic proton-related hadron radiation \citep{1993A&A...269...67M,2000NewA....5..377A,2003ApJ...586...79A}. According to the intensity of spectral emission lines, blazars are classified as BL Lacertae objects (BL Lacs) and Flat-spectrum Radio Quasars (FSRQs) \citep{2012MNRAS.420.2899G}. According to the synchrotron peaked frequency, blazars are also divided into four categories; Low Synchrotron Peaked, Intermediate Synchrotron Peaked, High Synchrotron Peaked, and Extremely High Synchrotron Peaked blazars \citep{2010ApJ...716...30A,2016ApJ...823L..26A}.

The light variability of blazars is very violent and usually exhibits irregular noise-like behavior. In the entire blazar family, there are only a few sources whose light curves change regularly, among which a very special type is the quasi-periodic oscillation (QPO). Periodic light variability is of great significance for searching the structure of binary black holes or studying the special dynamical behavior in the inner region of black holes. OJ 287 is the first blazar reported to have a QPO with a period time scale of 12 years in the optical V-band. This long-timescale QPO signal can be well explained in the binary black hole scenario \citep{1988ApJ...325..628S}. Subsequently, \cite{1992A&A...264...32K} analyzed the optical B-band light curve of OJ 287 for about 100 years and found a QPO with a time scale of ($11.6\pm 0.5$ years). Since then, some interesting QPO candidates have been reported in the optical, radio, X-ray, and gamma-ray bands: AO 0235+16, RE J1034+396, PKS 2155-304, 2XMM J123103.2+110648, J1359+4011, PG 1553+113, 1H 0707-495, PKS 0537-441, Mrk 766, PMN J0948+0022, PKS 0219-164, PKS 2155-304, PKS 0426-380, PKS 0301-243, OJ 287, PKS J2134-0153 , PKS J0805-0111, PKS 0521-36, and PKS 1510-089 \citep{2001A&A...377..396R, 2008Natur.455..369G, 2009A&A...506L..17L, 2013ApJ...776L..10L, 2013MNRAS.436L.114K, 2015ApJ...813L..41A, 2018ApJ...853..193Z, 2016ApJ...820...20S, 2017ApJ...849....9Z, 2017ApJ...849...42Z, 2017ApJ...847....7B, 2017ApJ...835..260Z, 2017ApJ...842...10Z, 2017ApJ...845...82Z, 2020MNRAS.499..653K, 2021MNRAS.506.3791R, 2021RAA....21...75R, 2021ApJ...919...58Z, 2023MNRAS.519.4893L}. 
These QPO candidates are mainly distributed on annual timescales, and most of them only appear in a single frequency band, with only a few sources (for example, PG 1553+113) appearing in two or more frequency bands. Long-time-scale QPOs have been more widely discussed in the supermassive binary black hole scenario, and these sources are also considered as possible candidates for supermassive binary black holes \citep{2008Natur.452..851V,2010MNRAS.402.2087V}. In addition, persistent jet precession \citep{2000A&A...360...57R,2004ApJ...615L...5R} and Lense-Thirring precession of accretion disks \citep{1998ApJ...492L..59S,2018MNRAS.474L..81L,2018ApJ...858...82Y} are also used in some scenarios to explain long-term QPO.
 
\cite{2018NatCo...9.4599Z} first reported a 34.5-day QPO in the gamma-ray light curve of the blazar PKS 2247-131, which is also the first monthly-time-scale transient QPO reported in an active galactic nucleus. This QPO appeared in the PKS 2247-131 gamma-ray flare state with six cycles and a time span of about 210 days. Its physical mechanism is attributed to a helical jet scenario with an enhanced radiation zone. Subsequently, many transient QPOs have also been reported one after another. The transient QPO of the high-redshift FSRQ B2 1520+31 in the gamma-ray band is 71 days, with seven cycles and a total duration of about 500 days \citep{2019MNRAS.484.5785G}. The FSRQ CTA 102 exhibits a transient QPO (quasi-periodic oscillation) event lasting approximately 7.6 days, simultaneously observed in both the gamma-ray and optical bands. This QPO event persists for about 60 days and undergoes eight cycles \citep{2020A&A...642A.129S}. There is a 47-day transient QPO in the gamma-ray variability of 3C 454 with nine cycles lasting 450 days, and a similar lower-confidence signal in the optical band \citep{2021MNRAS.501...50S}. The TeV blazar PKS 1510-089 exhibits a 3.6-day transient QPO with five cycles, lasting approximately 17 days. Additionally, it shows a 92-day transient QPO with seven cycles, lasting for 650 days \citep{2022MNRAS.510.3641R}. \cite{2022ApJ...938....8C} searche for monthly timescale QPO signals among 1525 highly variable sources given in the Fermi-LAT lightcurve library. They found a 31.3$\pm$1.8-day transient QPO in the gamma-ray light curve of TeV blazar S5 0716+714, with seven cycles lasting about 220 days. \cite{2023A&A...672A..86R} analyzed the light curves of the 35 brightest Fermi-LAT AGNs, including data from the start of the Fermi mission (August 2008) to April 2021, and energies from 100 MeV to 300 GeV. Two time-bin, 7 days and 30 days, were investigated. A search for quasi-periodic features was then performed using continuous wavelet transform, identifying 24 transient QPO candidates. Among them, there are 16 transient QPOs whose cycle times are greater than or equal to five. As shown in Table \ref{tab:tab1}, we list the information of the more credible transient QPOs that have been reported so far (including source name, 4FGL name, type, QPO time scale, QPO cycle number, and literature sources). The physical mechanism of these transient QPOs is could be attributed to hotspots rotating on the innermost stable circular orbit of the supermassive black hole \citep{1991A&A...246...21Z,2009ApJ...690..216G}, magnetic reconnection events of nearly equidistant magnetic islands within the jet \citep{2013RAA....13..705H,2018ApJ...854L..26S}, and helical motion of the enhanced emission region (or blob) inside the jet \citep{2015ApJ...805...91M,2017MNRAS.465..161S,2018NatCo...9.4599Z}.

The paper aims to systematically search and study for transient QPO events in the gamma-ray band for a flux-limited bright blazar sample. The frequency (period) we are interested in ranges from 0.0025-0.05 Hz (20-400 days). 
It is well known that periodic signals often hide within noise, making them difficult to detect and prone to false alarms. The confidence in their presence is also often overestimated.
Therefore, in this paper, we use more stringent criteria to screen and confirm transient QPOs. 
This work reports four new transient QPOs, also confirms four transient QPOs reported by previous works, and reevaluates the confidence level of signal detection. Furthermore, based on the search results of the flux-limited sample, we provide the first estimate of the incidence of transient QPO events. The possible physical origins of these transient QPO events are discussed.
The paper is organized as follows. Section \ref{section:2} describes the sample sources and the Fermi data processing procedure. Section \ref{section:3} describes the search process and methodology for transient QPOs. Section \ref{section:4} presents the analysis results of all transient QPOs. Section \ref{section:5} is a discussion, and Section \ref{section:6} is a summary.

\begin{table*}
   \centering
   \caption{Summary of the transient QPOs of individual blazars as reported in the literature.}
   \label{tab:tab1}
   \renewcommand\arraystretch{1.2}
  \setlength{\tabcolsep}{4mm}
   \begin{tabular}{lccccc} 
\hline
Source name  &  4FGL name   &  Source class   &  QPO time scale (days) & QPO cycle number &  Reference \\
\hline
PKS 2247-131   &  4FGL J2250.0-1250   &   BCU     &   34.5$\pm$1.5   &   6    &  a, b  \\
B2 1520+31     &  4FGL J1522.1+3144   &   FSRQ    &   179$\pm$42     &   6    &  b     \\
               &                      &           &   71$\pm$15      &   14   &  c, b  \\
               &                      &           &   39$\pm$11      &   17   &  b     \\
PKS 1510-089   &  4FGL J1512.8-0906   &   FSRQ    &   3.6$\pm$0.07   &   5    &  d     \\
               &                      &           &   92$\pm$1.2     &   7    &  d     \\
3C 454.3       &  4FGL J2253.9+1609   &   FSRQ    &   $47^{+0.97}_{-0.51}$    &   9   &  e     \\
3C 279         &  4FGL J1256.1-0547   &   FSRQ    &   39$\pm$1       &   -    &  f     \\
               &                      &           &   24$\pm$1       &   -    &  f     \\
               &                      &           &   101$\pm$27     &   6    &  b     \\
4C +01.02      &  4FGL J0108.6+0134   &   FSRQ    &   122$\pm$26     &   5    &  b     \\
PKS 0402-362   &  4FGL J0403.9-3605   &   FSRQ    &   122$\pm$42     &   5    &  b     \\
PKS 0426-380   &  4FGL J0428.6-3756   &   BL Lac  &   85$\pm$26      &   8    &  b     \\
PKS 0447-439   &  4FGL J0449.4-4350   &   BL Lac  &   111$\pm$42     &   7    &  b     \\
S5 0716+714    &  4FGL J0721.9+7120   &   BL Lac  &   324$\pm$77     &   5    &  b     \\
1H 1013+498    &  4FGL J1015.0+4926   &   BL Lac  &   52$\pm$12      &   12   &  b     \\
4C +21.35      &  4FGL J1224.9+2122   &   FSRQ    &   66$\pm$17      &   6    &  b     \\
PKS 1424-41    &  4FGL J1427.9-4206   &   FSRQ    &   90$\pm$22      &   5    &  b     \\
Mrk 501        &  4FGL J1653.8+3945   &   BL Lac  &   326$\pm$76     &   7    &  b     \\
CTA 102        &  4FGL J2232.6+1143   &   FSRQ    &   178$\pm$40     &   5    &  b     \\
               &                      &           &   $7.6^{+0.36}_{-0.25}$   &   8   &  g     \\
\hline
    \end{tabular} \\
   \footnotesize{Notes: Column (1) is the name of the object; Column (2) is the name of Fermi 4FGL; Column (3) is the type of source (BL Lac or FSRQ); Column (4) is the QPO time scale and error; Column (5) is the QPO cycle number; Column (6) is the reference: (a) \cite{2018NatCo...9.4599Z}, (b) Ren et al. (2023), (c) \cite{2019MNRAS.484.5785G}, (d) \cite{2022MNRAS.510.3641R}, (e) \cite{2021MNRAS.501...50S}, (f) \cite{2016AJ....151...54S}, (g) Sarkar et al. (2020).}
\end{table*}

\section{The sample of Fermi-LAT light curves} \label{section:2}
\subsection{Source Selection}
Fermi-LAT monitors a large number of extragalactic bright sources for a long time and adds them to the monitoring source list when the source flux exceeds the monitoring flux threshold of $1\times10^{-6}$~ph~cm $^{-2}$ s $^{-1}$ (\href{https://fermi.gsfc.nasa.gov/ssc/data/access/lat/msl\_lc/}{Monitored Source List}). As the mission progressed, the list continued to grow, and until now there are 197 sources. This catalog contains 173 Blazars of interest, including 19 Blazar Candidate of Uncertain type (BCU), 32 BL Lacs, and 122 FSRQs. To ensure the continuity and authenticity of the light curve, we add two conditions to further screen the samples: (1) Except for Fermi-LAT observations, these sources must be found by other telescopes to correspond to other wavelength bands. (2) The light curve for each source should have a sufficient number of data points. If more than 60\% of the data points are missing from the light curve, that source will be excluded from the sample. Based on the above two criteria, we finally identified 134 blazars as the research objects of this paper, which are given in Table \ref{tab:tab3}, consisting of 7 BCUs, 31 BL Lacs, and 95 FSRQs. For each source, the source name, 4FGL name, source type, coordinates, and redshift in J2000 are listed from left to right (the last three quantities available in the 4FGL catalog)

\subsection{Fermi-LAT Data Analysis}
The Fermi satellite was launched on June 11, 2008, and operates at an altitude of 565 km. The Fermi satellite carries two instruments, the Large Area Telescope (LAT) pair-conversion detector and the Gamma-ray Burst Monitor (GBM, \cite{2009ApJ...697.1071A}). Fermi-LAT is a high-energy gamma-ray telescope with an observation energy range from $\sim$20 MeV to 500 GeV, with a large field of view ($>$2 sr about 20\% of the all-sky). Since routine scientific observations began on August 4, 2008, the entire sky has been scanned every 3 hours. It provides an angular resolution of 5$^{\circ}$ per event below 100 MeV, downscaling to 2$^{\circ}$ at 300 GeV. The all-sky monitoring capabilities of the LAT provide us with nearly continuous observations of gamma-ray sources on different time scales.

We downloaded Fermi-LAT data for the Pass8 source class during MET 239557417-647393397 ($\sim$ 12.9 years). Based on the Fermi Science Toolkit $v11r05p3$ analysis was performed using standard analysis programs provided by the fermi-LAT collaboration. Select events belonging to the SOURCE class (evclass = 128, evtype = 3) with energies ranging from 100 MeV to 300 GeV, and exclude photons with zenith angles greater than $90^{\circ}$ (to reduce diffuse reflection from the Earth's atmosphere photon effects). We use the standard filter (DATA\_QUAL$>$0) \&\& (LAT\_CONFIG$==$1) to select a good time interval (GTI) to filter the data, and set the range of 15$^{\circ}$ centered on the point source as the region of interest (ROI). For the diffuse background model file, we choose the Galactic diffuse emission model $gll\_iem\_v07.fits$ (the model includes the sun, moon, Fermi bubble, and gamma-ray radiation generated by cosmic ray propagation in the Milky Way) and the isotropic model for point source $iso\_P8R3\_SOURCE\_V3\_v1.txt$ (the model used to describe the extragalactic gamma diffuse radiation background provided by the extragalactic sub detection source and some cosmic ray background that cannot be completely deducted by the instrument itself). Then, we use the script $make4FGLxml.py$ to create the source model XML file, which includes the location of the source and the best prediction of the spectral form. Perform a maximum likelihood analysis on the XML spectrum file using the $gtlike$ tool and the instrument response function (IRF) $P8R3\_SOURCE\_V3$ to obtain the spectrum of the source. In the 4FGL catalog, the final source spectrum is modeled using a log-parabola \citep{2020ApJS..247...33A}. We use the $gttsmap$ tool to perform a statistical test (TS) on the area around the target source, add a test point source model for each location, and use the formula TS=-2$(lnL0-lnL)$ to calculate the significant (L0 represents the maximum likelihood value of background model fitting, and represents the maximum likelihood value of fitting by adding test point sources in the background model). Through the above analysis and data reduction, we generated the flux data of the source. Subsequently, the light curves for each source were plotted, excluding the few data points with TS less than 9.

\section{Transient quasi-periodic oscillations identification} \label{section:3}
\begin{figure*}
\begin{minipage}[t]{1\textwidth}
\centering
\includegraphics[height=6.5cm,width=15cm]{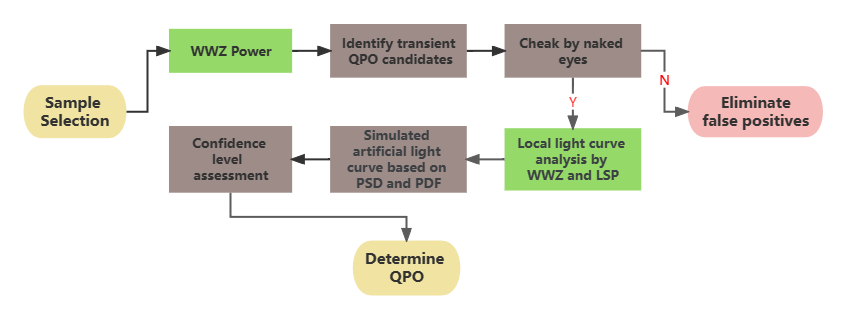}
\caption{Flowchart for Searching and Evaluating Transient QPOs.}
\label{Figure1}
\end{minipage}
\end{figure*}

Since the first case of transient QPO in AGN was reported in the gamma-ray emission of PKS 2247-131 \citep{2018NatCo...9.4599Z}, the study of transient QPO has attracted extensive attention and discussion. The identification of transient QPOs is more complex than that of long-standing QPOs. This paper utilized the WWZ technique, known for its sensitivity to transient QPO, to detect QPO signals. Subsequently, we verified and examined the signals by eye to account for possible false positives and harmonics. Confidence levels were assessed by Monte Carlo simulations. Here is a brief introduction to the specific analysis process, as shown in Figure \ref{Figure1}.

(1) Perform WWZ analysis on the global light curves of all samples, and determine the possible transient QPO on the wavelet power map.

(2) Determine the light curve of the transient QPO given by the WWZ method with the naked eye, and eliminate false positives.

(3) We independently analyze the light curves where transient QPOs occur using the WWZ and LSP methods, respectively. The artificial light curve is then generated from the power spectral density (PSD) and probability distribution function (PDF) of the original light curve to assess the confidence level of the periodic component.

\subsection{WWZ global search}
Compared with Fourier analysis, Wavelet Transform (WT) is more impressive in identifying transient QPO signals \citep{1995AJ....109.1889F}. WT evaluates the signal over a certain time range, searching for frequency components in each time segment separately to identify local properties in the time and frequency domains. Also, the WT time range width can be scaled to generate multiple resolutions in time and frequency by traversing different time scales. Therefore, it is sensitive to the local attributes and global attributes of time series data and can detect possible local transient periodic fluctuations and parameter changes in the signal. The continuous wavelet transform of a time series $x(t)$ is defined as
\begin{equation}
\begin{aligned}
W(\omega,\tau;x(t)) = \omega^{1/2}\int x(t)f^{\ast}(\omega(t-\tau))dt=\omega^{-1/2} \\
\int x(\omega^{-1}z+\tau)f^{\ast}(z)dz,
\label{eq:LebsequeIp1}
\end{aligned}
\end{equation}
where $f^{\ast}$ is the complex conjugate of $f$, the function $f(z)$ is the mother wavelet, $\omega$ is the scaling factor and $\tau$ is the time shift.
For a discrete dataset of observations x\{$t_\alpha$, $\alpha$=1, 2, ..., N\}, the wavelet transform can be written as the discrete wavelet transform (DWT)
\begin{equation}
W(\omega,\tau;x(t)) = \sqrt{\omega}\sum\limits^N_{\alpha=1} x(t_a)f^{\ast}(\omega(t_a-\tau)),
\label{eq:LebsequeIp2}
\end{equation}
introduce the mother wavelet function the Morlet wavelet $f(z)$
\begin{equation}
f(z)=e^{iz-cz^2}=e^{i\omega(t-\tau)-c\omega^2(t-\tau)^2},
\label{eq:LebsequeIp3}
\end{equation}
replace the number of data points N with the effective number
\begin{equation}
N_{eff}=\frac{(\sum\omega_{\alpha})^{2}}{\sum\omega_{\alpha}^{2}}=\frac{[\sum e^{-c\omega^2(t-\tau)^2}]^2}{\sum e^{-2c\omega^2(t-\tau)^2}}.
\label{eq:LebsequeIp4}
\end{equation}
Then Foster defines the Weighted wavelet transform (WWT),
\begin{equation}
\begin{aligned}
WWT=\frac{(N_{eff}-1)V_{y}}{2V_{x}}  \\
V_{x}=\frac{\sum\limits_\alpha\omega_{\alpha}x^{2}t_{\alpha}}{\sum\limits_\beta\omega_{\beta}}-
\left [\frac{\sum\limits_\alpha\omega_{\alpha}x t_{\alpha}}{\sum\limits_\beta\omega_{\beta}}\right]^{2} \\
V_{y}=\frac{\sum\limits_\alpha\omega_{\alpha}y^{2}t_{\alpha}}{\sum\limits_\beta\omega_{\beta}}-
\left [ \frac{\sum\limits_\alpha\omega_{\alpha}y t_{\alpha}}{\sum\limits_\beta\omega_{\beta}} \right] ^2,
\label{eq:LebsequeIp6}
\end{aligned}
\end{equation}
where $V_{x}$ and $V_{y}$ is the weighted variation of the data and model function, respectively \citep{1996AJ....112.1709F}. For sine or cosine signals, WWT tends to peak at lower frequencies. At lower frequencies, we have a wider "window" and thus effectively sample more data points (with a larger effective number $N_{eff}$), which makes the WWT increase as $\omega$ decreases and the data fit worse. The Z-statistic is a test statistic that is less sensitive to the amount of valid data, which will give a better estimate of the frequency of important peaks. Also known as Weighted Wavelet Z-transform \citep{1996AJ....112.1709F}.
\begin{equation}
Z=\frac{(N_{eff}-3)V_{y}}{2(V_{x}-V_{y})},
\label{eq:LebsequeIp7}
\end{equation}
We implement the WWZ algorithm using the Python code provided by \cite{2017zndo....375648A} (the author declares that the code is derived from the Fortran code provided by \cite{2004JAVSO..32...41T}).
In this work, we perform a WWZ global analysis of all light curves in 134 samples. To more accurately search for potential transient QPO signals in long-term light curves, we determined that the frequency range of interest for WWZ analysis was 0.0025-0.05 $d^{-1}$ (20-400 days). In the contour power map given by the WWZ method, we identified 38 high-power transient QPOs and their appearance time ranges (MJD) in the 35 light curves. To rule out false positive signals, we visually traversed the local light curves of these 38 transient QPOs to confirm the authenticity of the signals. We identified three screening conditions: (1) The number of transient QPO cycles is greater than or equal to five \citep{2016MNRAS.461.3145V}. (2) The number of missing data points in a single cycle cannot exceed 60\%. (3) The transient QPO signal can be recognized by the naked eyes. Only when the light curve of a transient QPO satisfies the above three conditions, we can pay attention to it and add it to the transient QPO catalog (Table \ref{tab:tab2}) for further analysis. Eight transient QPO signals were preliminarily identified in the six sources (4C +01.02, PKS 0336-01, PKS 0402-362, PKS 0537-441, PKS 1424-41, and PKS 1510-089). It is worth noting that there are two transient QPO signals in the light curves of PKS 0537-441 and PKS 1424-41.

Quasi-periodic oscillations in blazar light curves are known to be very rare and thus difficult to detect and identify. Compared with previous work, we have given a more stringent screening condition, and there is no doubt that the results we present are very interesting. Next, we analyze and discuss these eight transient QPOs.

\subsection{Transient QPO determination}
For the eight transient QPOs signals identified above, we extracted the light curve during QPO from the whole light curve according to the QPO time range given by wavelet analysis for local research. For each light curve, the WWZ frequency range of interest is reset according to the period of each transient QPO, and finally, the power contour profile is drawn according to the power at each period position given by the WWZ method. Then, we introduce the LSP method to independently test the results given by the WWZ method \citep{1976Ap&SS..39..447L,1982ApJ...263..835S,2018ApJS..236...16V}. The LSP method is a well-known algorithm for detecting and characterizing the periodicity of time series and is widely used in time-domain astronomy research. This method can efficiently calculate the Fourier power spectrum of non-uniform sampling data and obtain possible periodic oscillation components. The following is a brief introduction to this method, assuming a data set $g(t)$ consisting of N equally spaced data points, according to the power spectrum, its classic periodogram can be defined as
\begin{equation}
P_s(f)=\frac{1}{N}\left| \sum^N_{n=1}g_n e^{-2\pi i\beta_n}\right|^2.
\label{eq:LebsequeIp8}
\end{equation}
When the classical periodogram is applied to uniformly sampled Gaussian white noise, the periodogram presents a chi-square distribution, and the periodogram can be used to distinguish periodic components from the non-periodic components.
The uniformly sampled classical periodogram can be generalized to non-uniformly sampled datasets by a simple change
\begin{equation}
P(f)=\frac{1}{N}\left[ \left(\sum^N_{n=1}g_n cos(2\pi f t_n)\right)^2+\left(\sum^N_{n=1}g_n sin(2\pi f t_n)\right)^2 \right].
\label{eq:LebsequeIp9}
\end{equation}
The statistical characteristics of this form of periodogram are more complex, and to a certain extent, it can identify potential periodic signals in non-uniform sampling data, but its periodogram cannot clearly distinguish periodic components from non-periodic components. Scargle addresses this problem by considering a generalized form of the periodogram \citep{1982ApJ...263..835S}
\begin{equation}
\begin{aligned}
P(f)=\frac{A^2}{2}\left(\sum_n g_n cos(2\pi f (t_n-\tau))\right)^2+ \\
\frac{B^2}{2} \left(\sum_n g_n sin(2\pi f (t_n-\tau))\right)^2,
\label{eq:LebsequeIp10}
\end{aligned}
\end{equation}
where $A$, $B$, and $\tau$ are arbitrary functions of frequency $f$ and observation time ${t_i}$. The Scargle-corrected periodogram fits the sinusoidal function model to the data of each frequency and constructs the periodogram according to the goodness of fit. Lomb also discusses this model in some depth \citep{1976Ap&SS..39..447L}, so it ends up being called Lomb-Scargle periodogram, whose generalized form is
\begin{equation}
\begin{aligned}
P_{LS}(f)=\frac{1}{2}\left( \frac{[\sum_n g_n cos(2\pi f (t_n-\tau))]^2}{\sum_n cos^2(2\pi f (t_n-\tau))} \right)  \\
+ \frac{1}{2}\left(\frac{[\sum_n g_n sin(2\pi f (t_n-\tau))]^2}{\sum_n sin^2(2\pi f (t_n-\tau))} \right),
\label{eq:LebsequeIp11}
\end{aligned}
\end{equation}
where the time phase correction $\tau$ is
\begin{equation}
\tau=\frac{1}{4\pi f}tan^{-1}\frac{\sum_n sin(4\pi f t_n)}{\sum_n cos(4\pi f t_n)}.
\label{eq:LebsequeIp12}
\end{equation}
Similar to the WWZ analysis, we analyzed the local light curves of the eight transient QPOs identified above using the LSP method to identify potential periodic components in the light curves and finally draw LSP power contours. Next, we performed Monte Carlo simulations based on red noise to determine the confidence level of the transient QPO determined by the WWZ and LSP methods, and to determine the importance of the periodic component in the light curve. We simulate the artificial light curve according to the method given by \cite{2013MNRAS.433..907E}, which uses a smooth bending power-law model plus Poisson noise to model the PSD of the original light curve, and at the same time, uses gamma and lognormal distributions to model PDF modeling of the original light curve. Then simulate the light curve of the same PSD and PDF distribution, here we use the Python program given by \cite{2016ascl.soft02012C} to implement the simulation algorithm. Finally, we separately simulated 10000 artificial light curves with the same PSD and PDF distributions as the original light curves for the local light curves of each transient QPO. In the confidence evaluation stage, first, calculate the WWZ power and LSP power of each artificial dimming curve, then calculate 99\%, 99.7\%, and 99.994\% confidence power, and finally draw 99\%, 99.7\%, and 99.994\% confidence contour lines in the frequency range of interest for each light curve in the WWZ power map and LSP power map, respectively.

\section{Results} \label{section:4}
In this section, we present the analysis results of the WWZ and LSP methods for all transient QPOs. Figure \ref{Figure2}, Figure \ref{Figure3}, Figure \ref{Figure4}, Figure \ref{Figure5}, Figure \ref{Figure6}, and Figure \ref{Figure7} respectively give 4C+01.02, PKS 0336-01, PKS 0402-362, PKS 0537-441, TXS 1219+285, PKS 1424-41 and PKS 1510-089 analytical results of transient QPO in gamma-ray. Except for QPOs of 253 days in 4C +01.02 and 103 days in PKS 0402-362, the confidence level of the other 6 QPOs exceeds 3$\sigma$. And 91.8 days in PKS 1510-089 and 55 days and 54.7 days in PKS 0537-441, the confidence levels of these three transient QPOs are all greater than or equal to 4$\sigma$. The time scale of QPOs is all within one year, the longest being the 341-day QPO in PKS 1424-41, and the shortest being the 54.7-day QPO in PKS 0537-441. Below we introduce the transient QPO in each source in detail.
\begin{table*}
   \centering
   \caption{List of transient QPO candidates and their associated confidence levels.}
   \label{tab:tab2}
   \renewcommand\arraystretch{1.2}
  \setlength{\tabcolsep}{3mm}
   \begin{tabular}{lcccccc} 
\hline
Source name & Source class & Time span (MJD) & WWZ period (days) & LSP period (days) & Significance  & QPO cycle number \\
\hline
4C +01.02      &  FSRQ    &   56422-58000   &   252.2 $\pm$ 20.9   &   253.2 $\pm$ 20.0   &   $\sim$ 2.7 $\sigma$        &   6   \\
PKS 0336-01    &  FSRQ    &   57020-57600   &   94.5 $\pm$ 7.2   &   94.6 $\pm$ 6.8  &   $\sim$ 3.4 $\sigma$   &   6   \\
PKS 0402-362   &  FSRQ    &   56817-57363   &   103.5 $\pm$ 7.9   &   103.9 $\pm$ 6.7  &   $\sim$ 2.6 $\sigma$        &   5   \\
PKS 0537-441   &  BL Lac  &   56803-57183   &   55.1 $\pm$ 3.1   &   55.0 $\pm$ 3.3   &   $\sim$ 4.1 $\sigma$   &   7   \\
               &          &   57636-58036   &   54.6 $\pm$ 2.9   &   54.7 $\pm$ 3.3   &   $\sim$ 4.0 $\sigma$   &   7   \\
PKS 1424-41    &  FSRQ    &   56998-57331   &   57.2 $\pm$ 4.1   &   57.1 $\pm$ 4.5   &   $\sim$ 3.2 $\sigma$   &   6   \\
               &          &   56611-58644   &   336 $\pm$ 37.3   &   341 $\pm$ 25.8   &   $\sim$ 3.7 $\sigma$   &   6   \\
PKS 1510-089   &  FSRQ    &   58197-58955   &   91.8 $\pm$ 5.3   &   92.3 $\pm$ 5.2  &   $\sim$ 4.3 $\sigma$   &   8   \\
\hline
    \end{tabular} \\
   \footnotesize{Notes: Column (1) is the name of the object; Column (2) is the type of source; Column (3) is the time span of the transient QPO; Column (4) shows the QPO and error given by the WWZ method; Column (5) shows the QPO and error given by the LSP method; Column (6) is the confidence level of the peak signal (take the maximum significance given by the previous two methods); and Column (7) is the QPO cycle number.}
\end{table*}

\subsection{4C +01.02}
\begin{figure*}
\begin{minipage}[t]{1\textwidth}
\centering
\includegraphics[height=7cm,width=14cm]{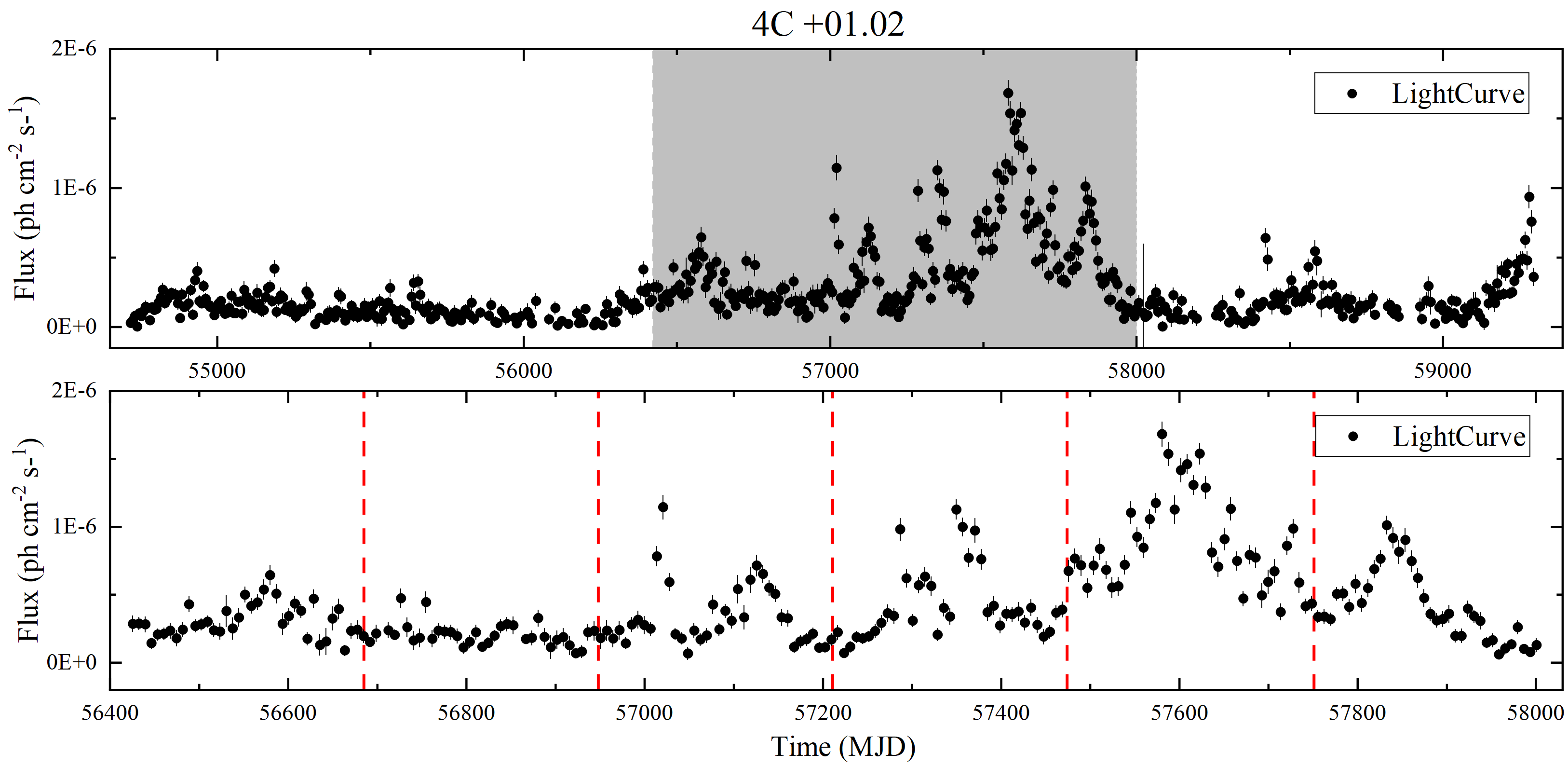}
\end{minipage}
\begin{minipage}[t]{1\textwidth}
\includegraphics[height=6cm,width=7cm]{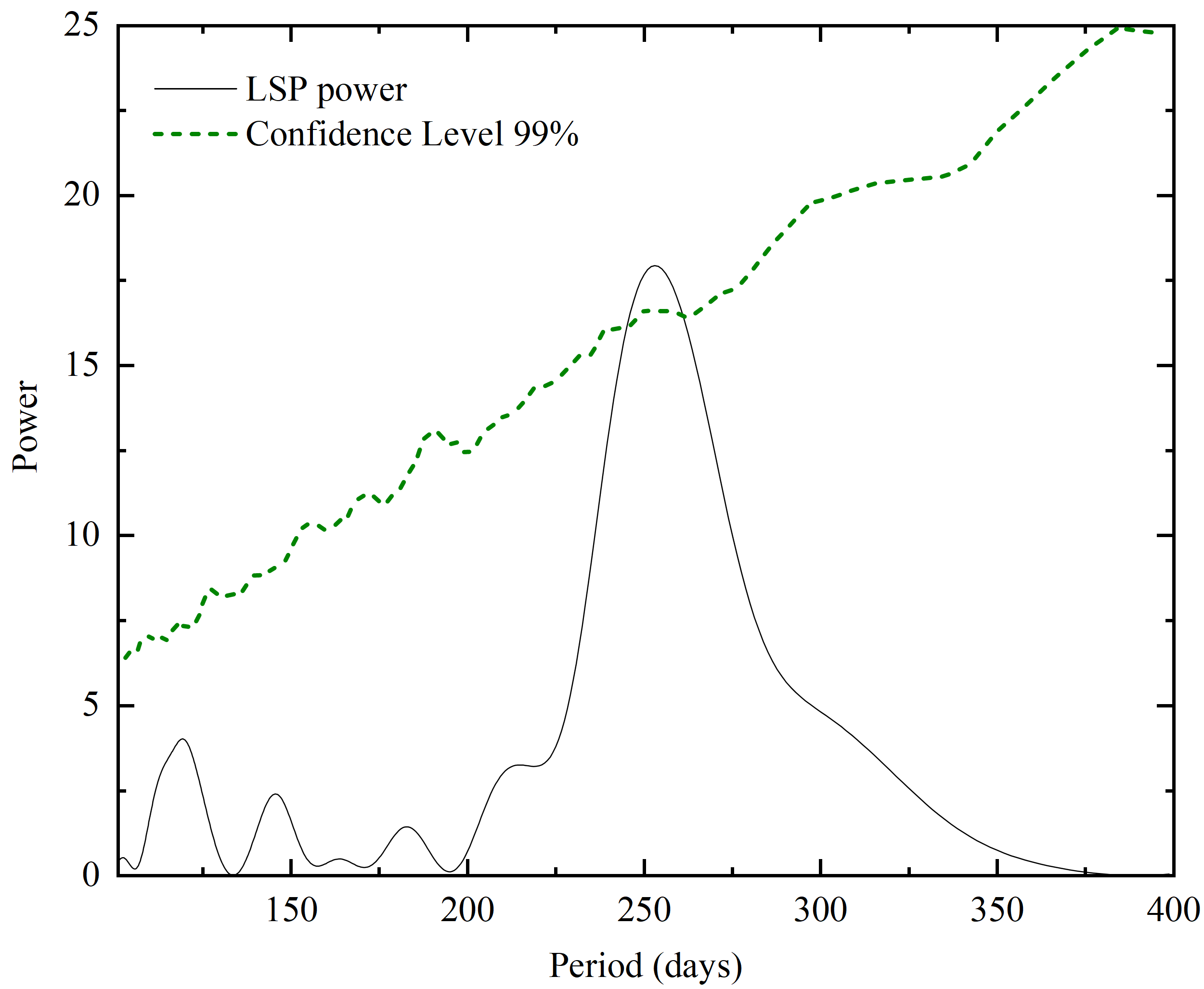}
\includegraphics[height=6.1cm,width=10.6cm]{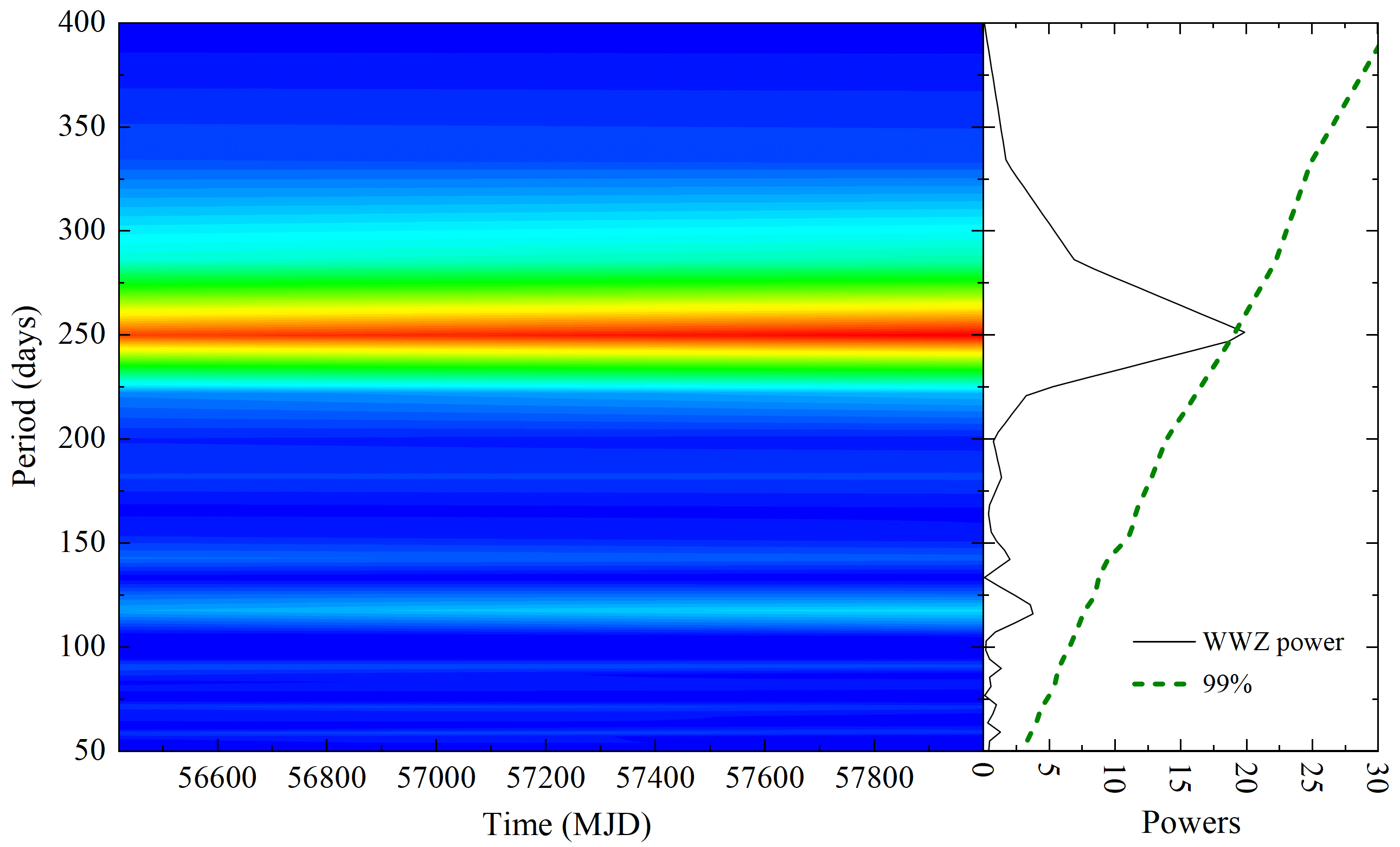}
\end{minipage}
\caption{Top panel: The graph above shows the weekly binned light curve for 4C +01.02, with MJD on the horizontal axis, flux and error on the vertical axis, and the light curve where the transient QPO phase occurs is shaded in gray. The figure below shows a magnified view of the gray shaded area, where the red dotted line is used to describe the cycle number of QPO. Bottom panel: The left panel shows the LSP power spectrum and significance evaluation results, where the black solid line is the LSP power and the green dashed line is the 99\% confidence level contour. The figure on the right shows the wavelet power spectrum and significance evaluation results, where the black solid line is the WWZ power, and the green dotted line is the 99\% confidence level contour.}
\label{Figure2}
\end{figure*}

4C +01.02 is a high-redshift FSRQ (z=2.099), which has GeV radiation observed by Fermi-LAT in 2013. Multi-wavelength studies have shown that the light variability of this source is very active, and the maximum flux of gamma-ray even reaches 61 times of the basic flux. The one zone leptonic model can well reproduce the broadband energy spectrum distribution, but its different flare states need to consider the contribution of BLR and IR torus seed photons to the EC process \citep{2022MNRAS.514.4259M}.
Figure \ref{Figure2} shows the light curve of 4C +01.02 and the analysis results of the local transient period. It can be seen intuitively from the figure that there is a QPO MJD 56422-58000, with the period of 253.2 $\pm$ 20.0 days and six cycles (The uncertainty is estimated as the half width at half maximum of the Gaussian function fitting the power peak. As shown in Figure \ref{Figure2}, the red dotted line gives the number of cycles). Similar QPOs have been reported in the related literature, and their analysis results showed that there were 286 and 123 days of QPO between MJD 56900-58000 and MJD 57300-57900, respectively, with four and five cycles, respectively \citep{2023A&A...672A..86R}. Among them, the 286-day QPO is relatively close to the 253-day result given in this paper. It is worth noting that the results given in the relevant literature are based on 30-day binning data, while our 253-day results are based on 7-day binning data. Overall, these results all point to an interesting fact that there is a transient QPO behavior in FSRQ 4C +01.02, and this periodic light variability may originate from some special activities of the black hole.

\subsection{PKS 0336-01}
PKS 0336-01 is an FSRQ with a redshift of 0.852, and its radio, optical, infrared, and gamma-ray band light variability all have obvious flares. As shown in Figure \ref{Figure3} for the PKS 0336-01 gamma-ray local light curve, we first report a 94.6 $\pm$ 6.8 day transient QPO with six cycles between MJD 57020-57600. Figure \ref{Figure3} also shows the analysis results of the local light curve WWZ and LSP methods and the confidence level contours. The confidence level of the periodic modulation signal in the light curve exceeds 3$\sigma$. Therefore, this is also a more plausible example of a transient QPO.
\begin{figure*}
\begin{minipage}[t]{1\textwidth}
\centering
\includegraphics[height=7cm,width=14cm]{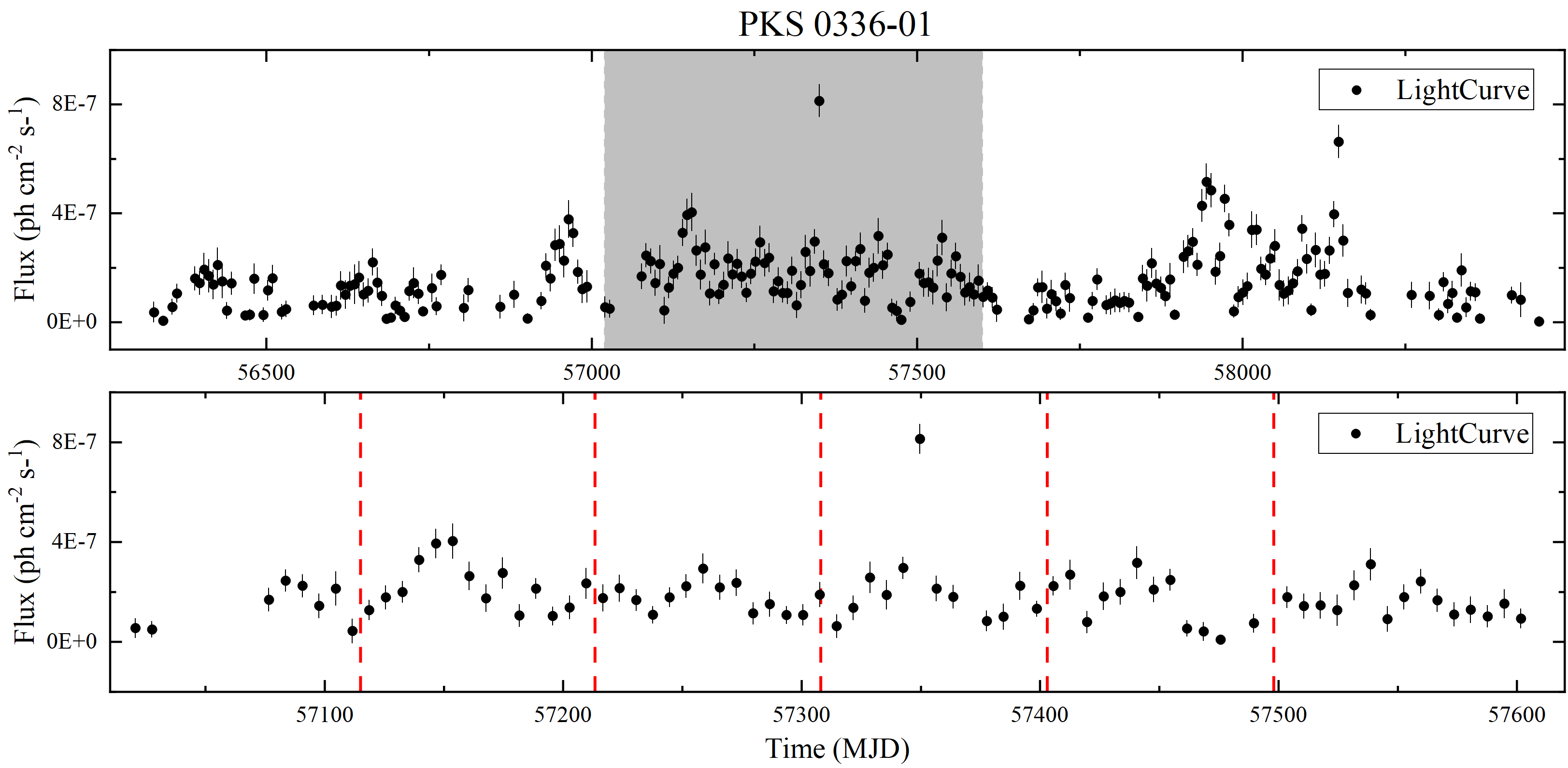}
\end{minipage}
\begin{minipage}[t]{1\textwidth}
\includegraphics[height=6cm,width=7cm]{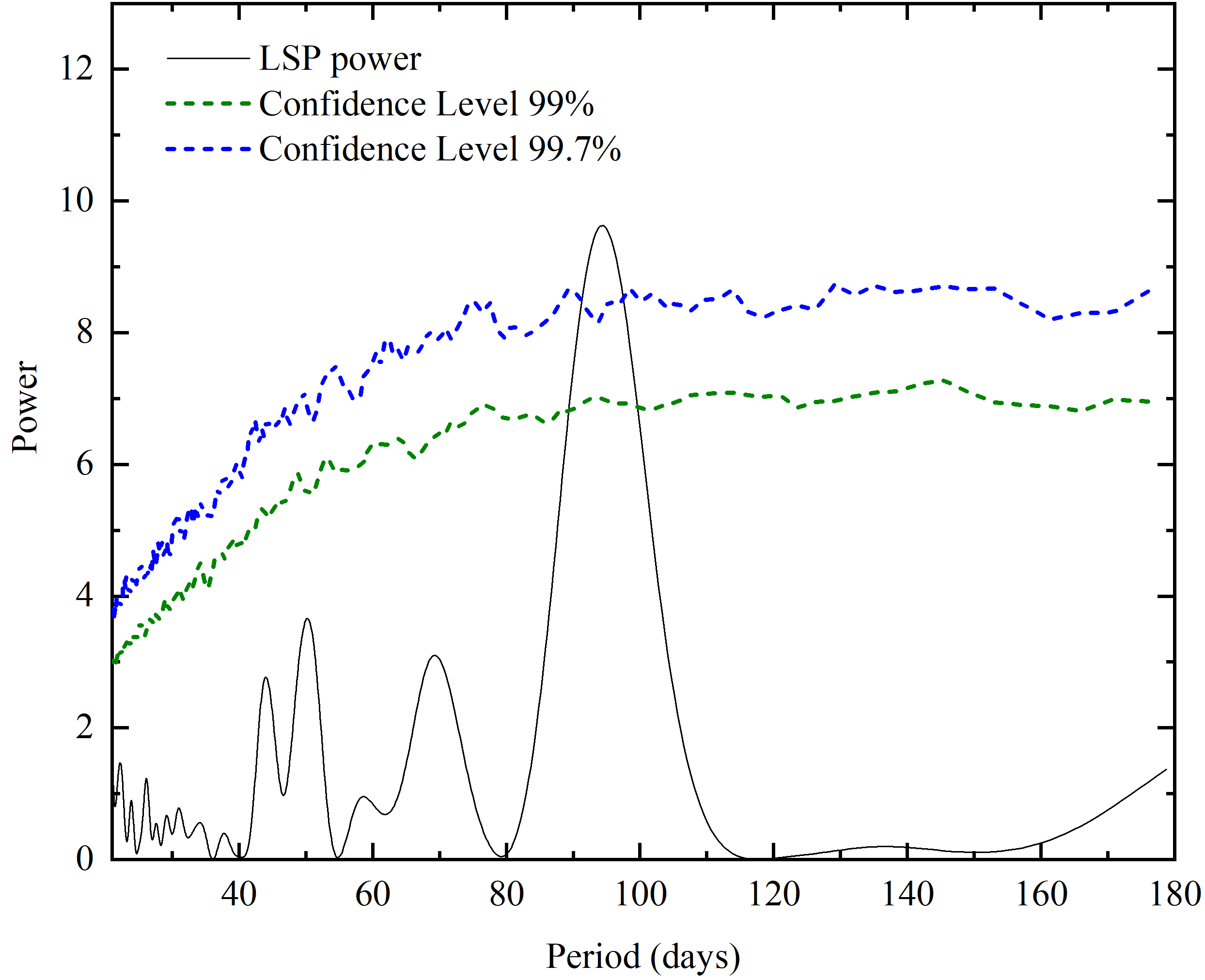}
\includegraphics[height=6.1cm,width=10.6cm]{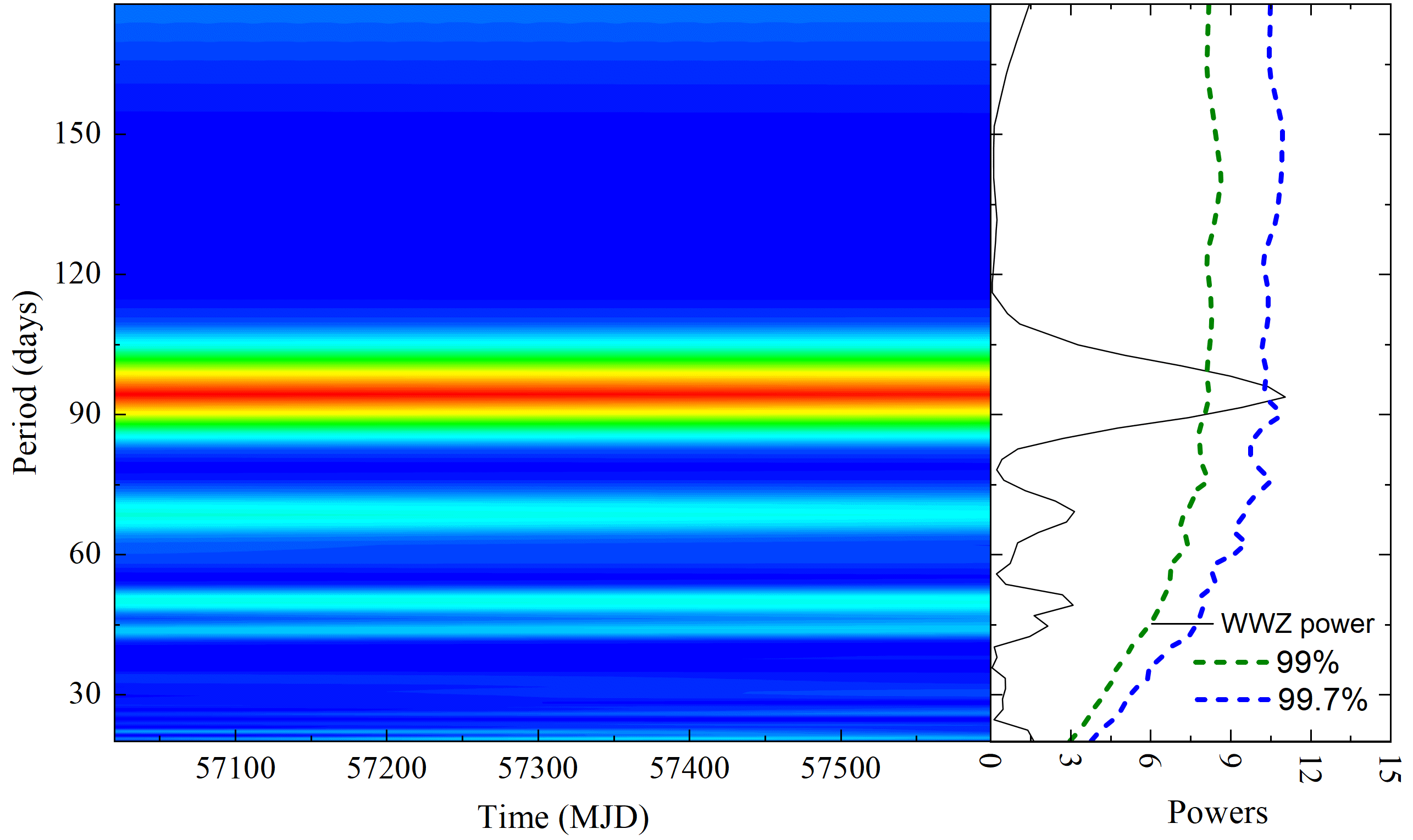}
\end{minipage}
\caption{Same description as Figure \ref{Figure2} for PKS 0336-01 (The blue dotted lines in the bottom panel represent the 99.7\% confidence level contours). \label{Figure3}}
\label{Figure3}
\end{figure*}

\subsection{PKS 0402-362}
\begin{figure*}
\begin{minipage}[t]{1\textwidth}
\centering
\includegraphics[height=7cm,width=14cm]{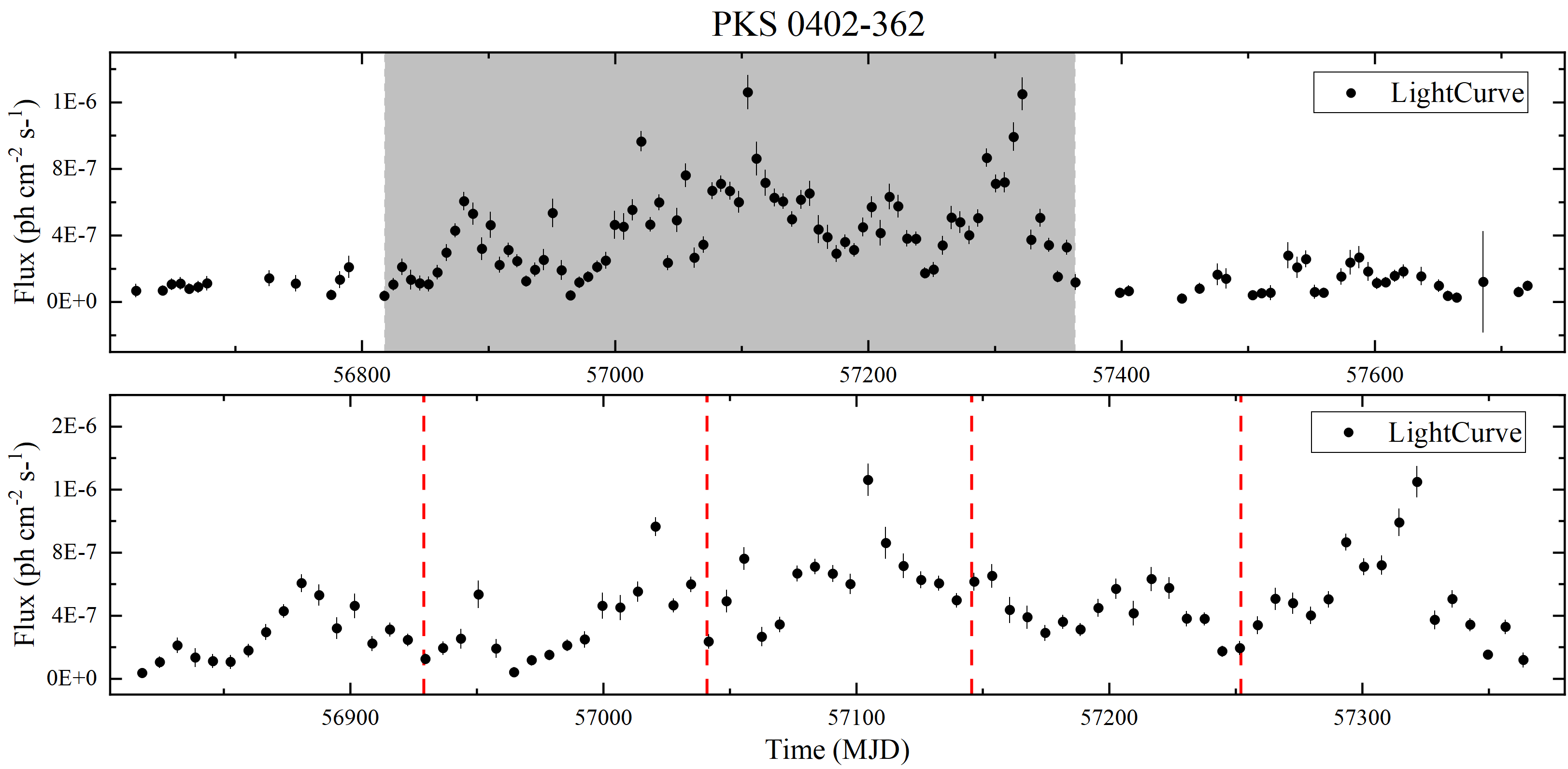}
\end{minipage}
\begin{minipage}[t]{1\textwidth}
\includegraphics[height=6cm,width=7cm]{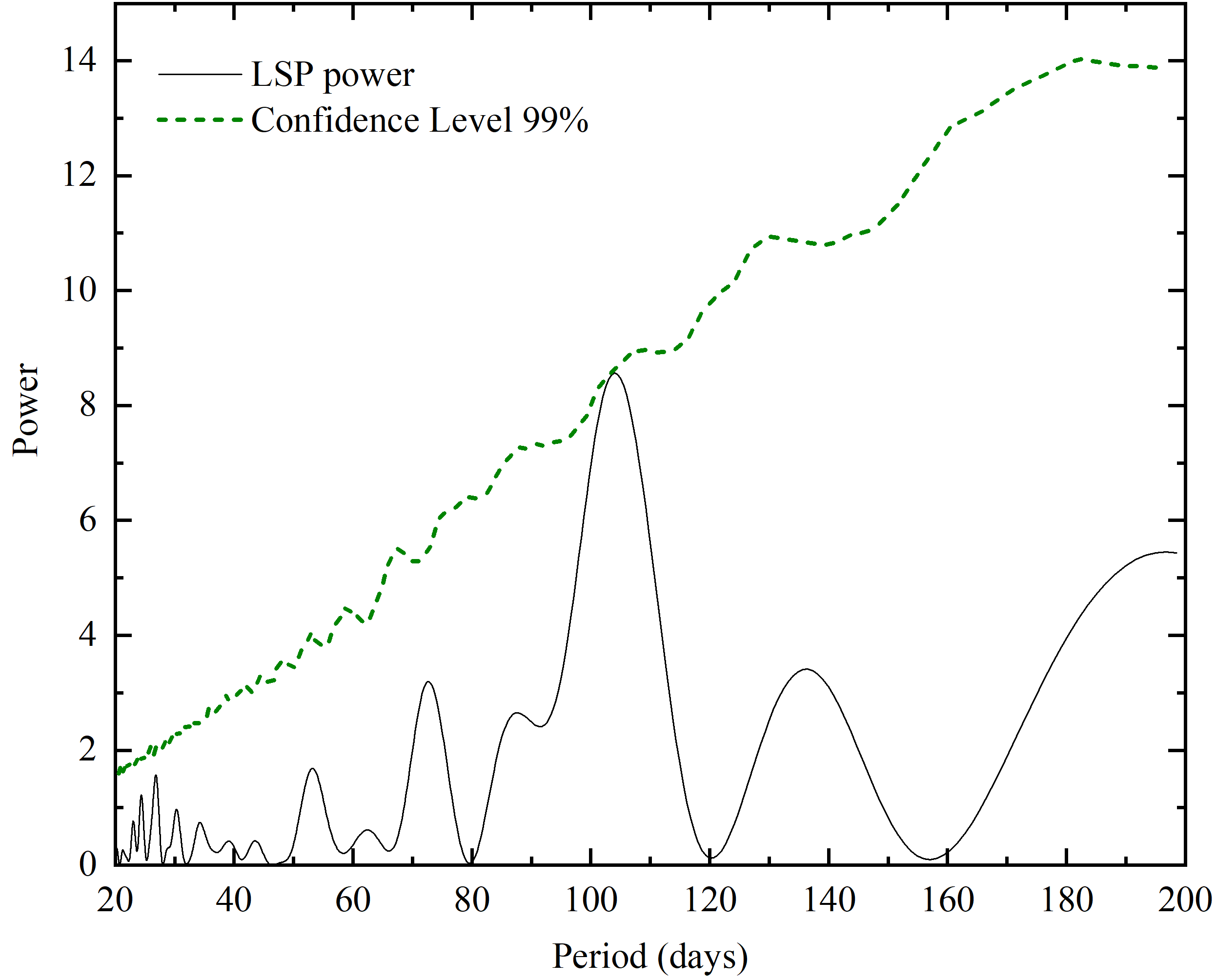}
\includegraphics[height=6.2cm,width=10.6cm]{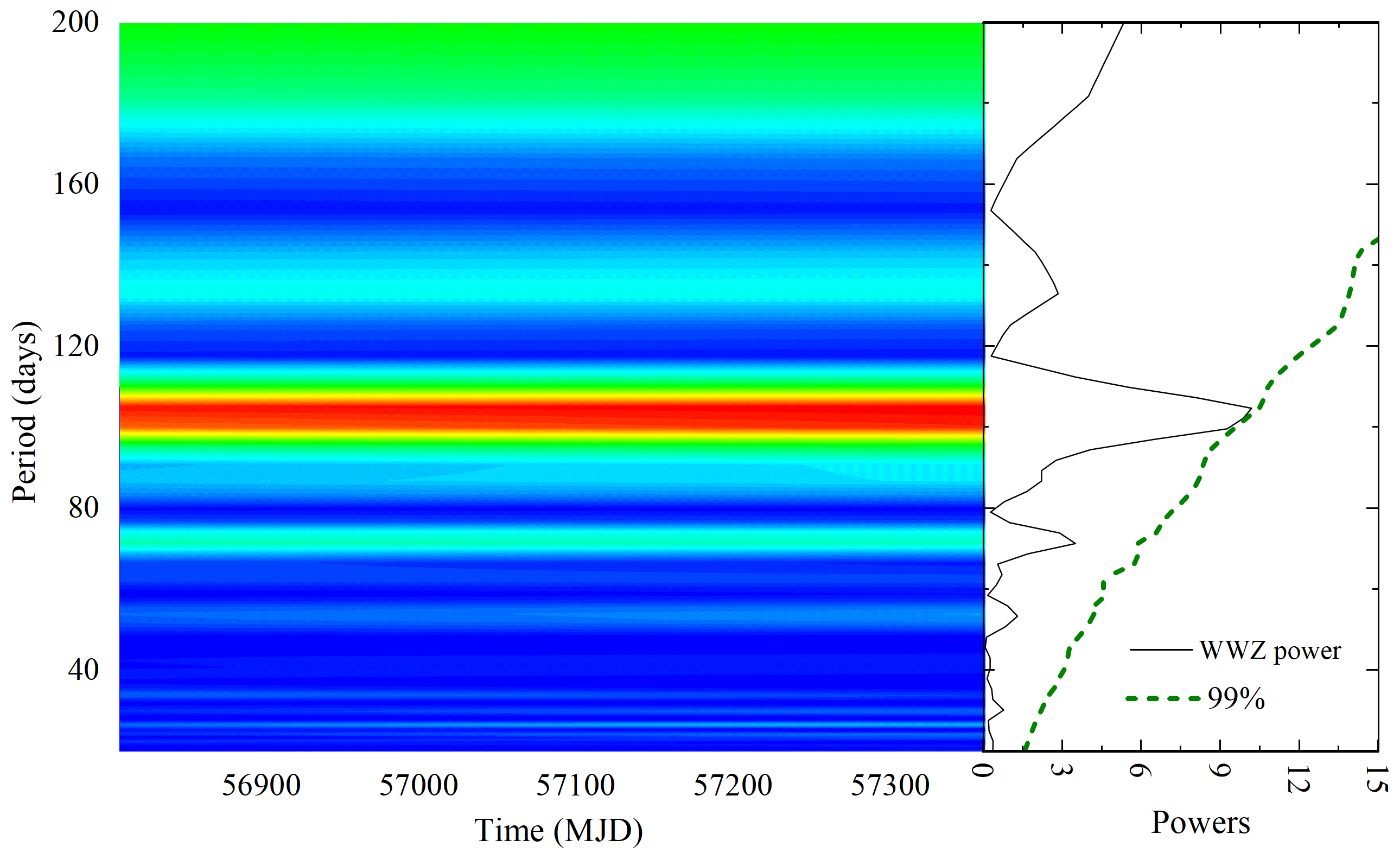}
\end{minipage}
\caption{Same description as Figure \ref{Figure2} for PKS 0402-362.}
\label{Figure4}
\end{figure*}

PKS 0402-362 is an FSRQ with a redshift of 1.5. \cite{2023MNRAS.521.3451D} studied the long-term light curve of gamma-ray for about 13 years. They found that most of the gamma-ray flare peaks had an asymmetrical profile, suggesting slow particle cooling times or changes in the Doppler factor were the main cause of these flares. There is no time lag between optical-IR and gamma-ray, consistent with a simple one-zone emission model. Among other things, they found that the light emission at different flux phases is dominated by the hot disk, suggesting that it is a good source for examining disk-jet coupling. The light curve is shown in Figure \ref{Figure4}, and we report a 103.5 $\pm$ 7.9-day transient QPO between MJD 56817-57363, which experienced five cycles. There are similar reports in the related literature \citep{2023A&A...672A..86R}. They detected 221-day and 112-day QPO in the 30-day binned data and 7-day binned data, respectively, and believed that the 112-day QPO may be a harmonic frequency of 221 days.

\subsection{PKS 0537-441}
\begin{figure*}
\begin{minipage}[t]{1\textwidth}
\centering
\includegraphics[height=9cm,width=14cm]{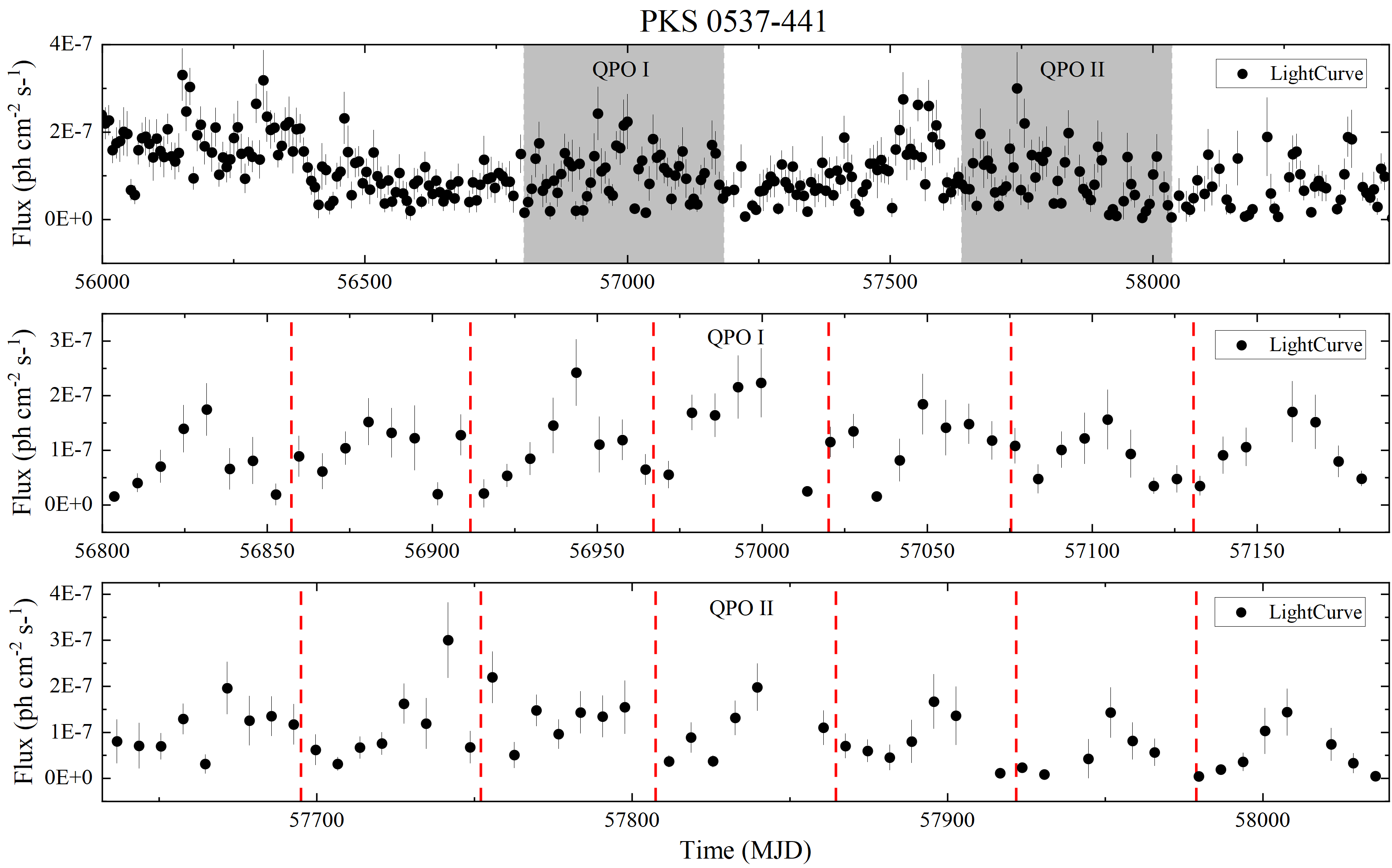}
\end{minipage}
\begin{minipage}[t]{1\textwidth}
\includegraphics[height=6cm,width=7cm]{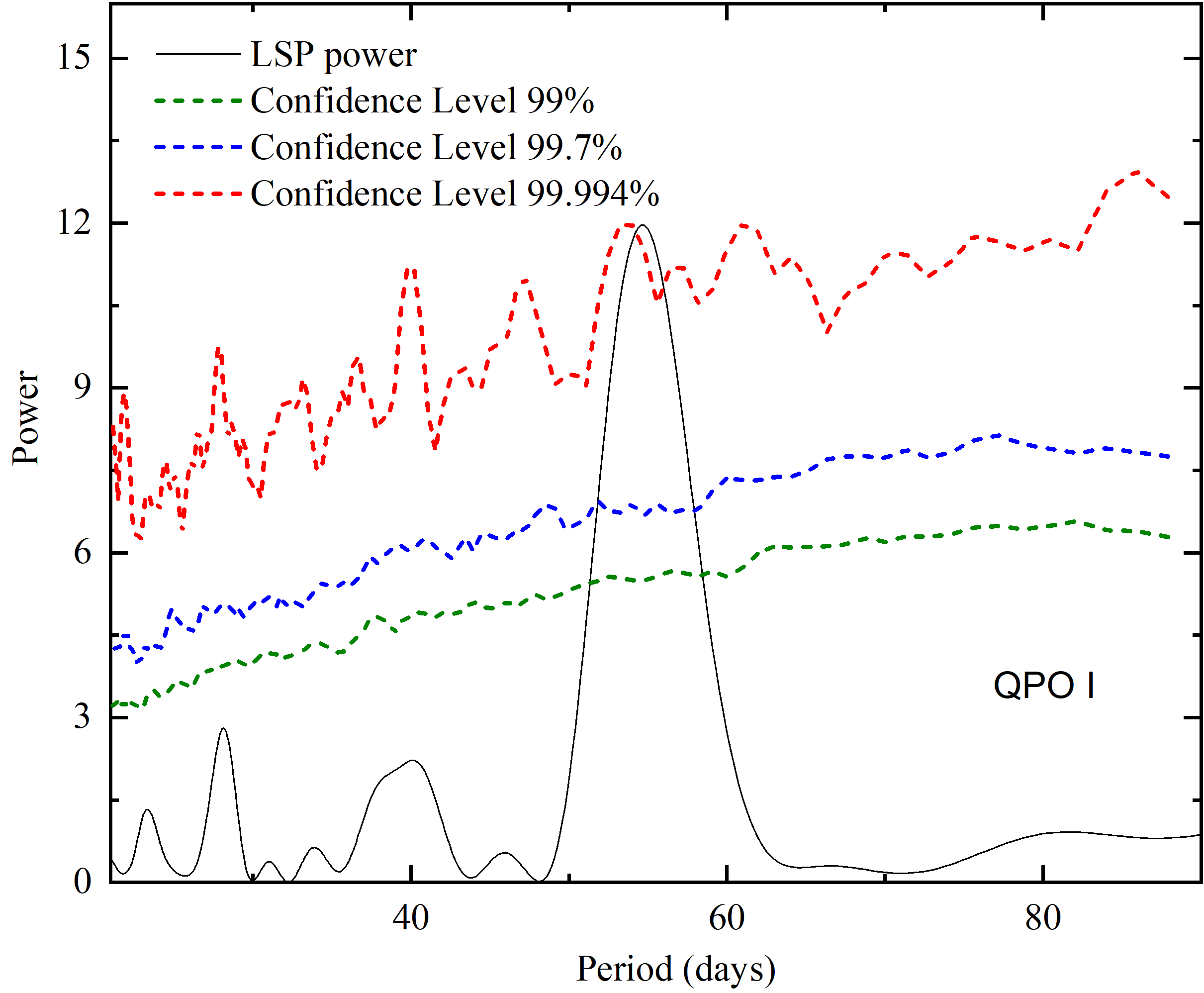}
\includegraphics[height=6.2cm,width=10.6cm]{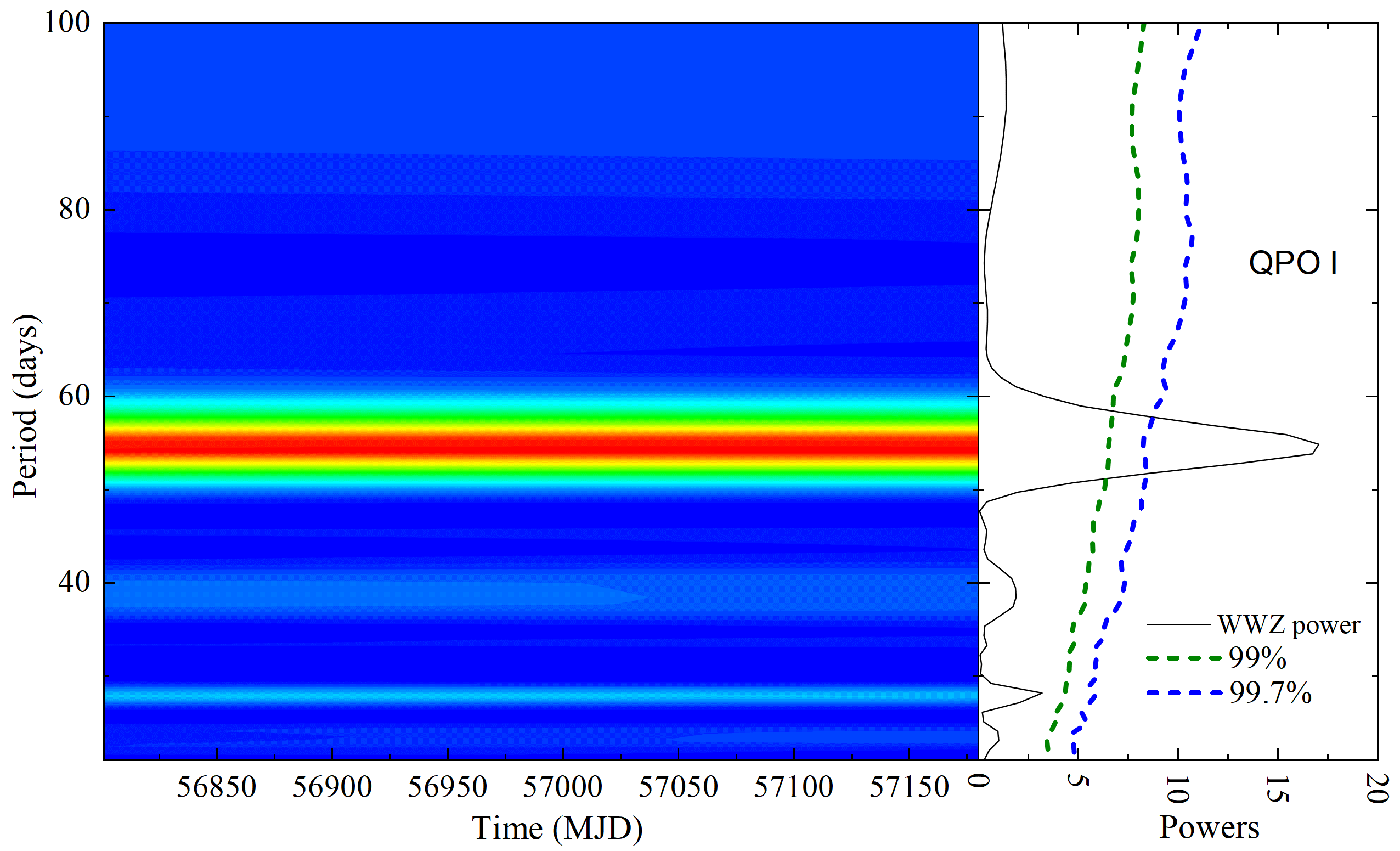}
\end{minipage}
\begin{minipage}[t]{1\textwidth}
\includegraphics[height=6cm,width=7cm]{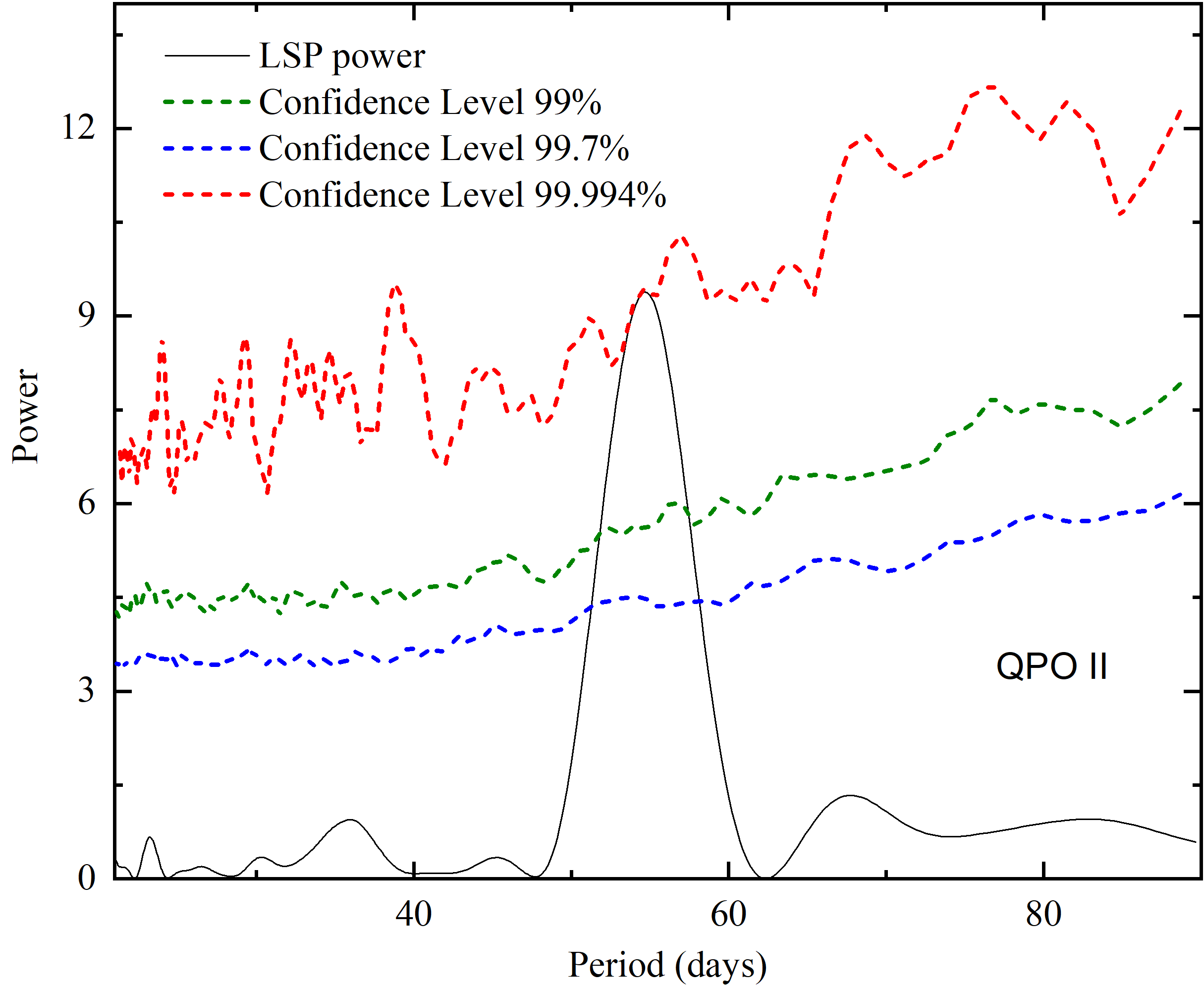}
\includegraphics[height=6.2cm,width=10.6cm]{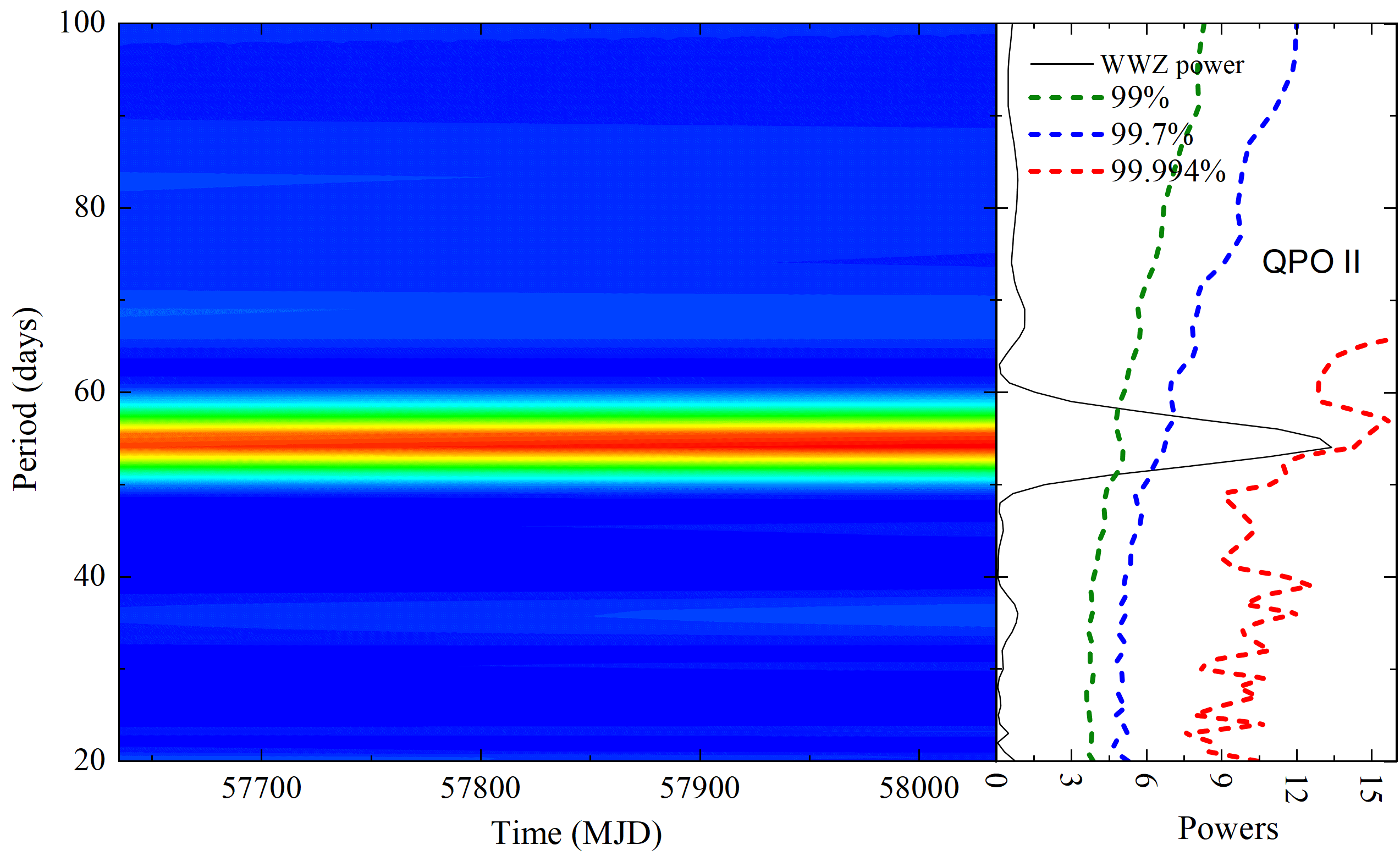}
\end{minipage}
\caption{Same description as Figure \ref{Figure2} for PKS 0537-441 (The blue and red dashed lines in the bottom panel represent the 99.7\% and 99.99\% confidence level contours, respectively).}
\label{Figure5}
\end{figure*}

\begin{figure*}
\begin{minipage}[t]{1\textwidth}
\centering
\includegraphics[height=9cm,width=14cm]{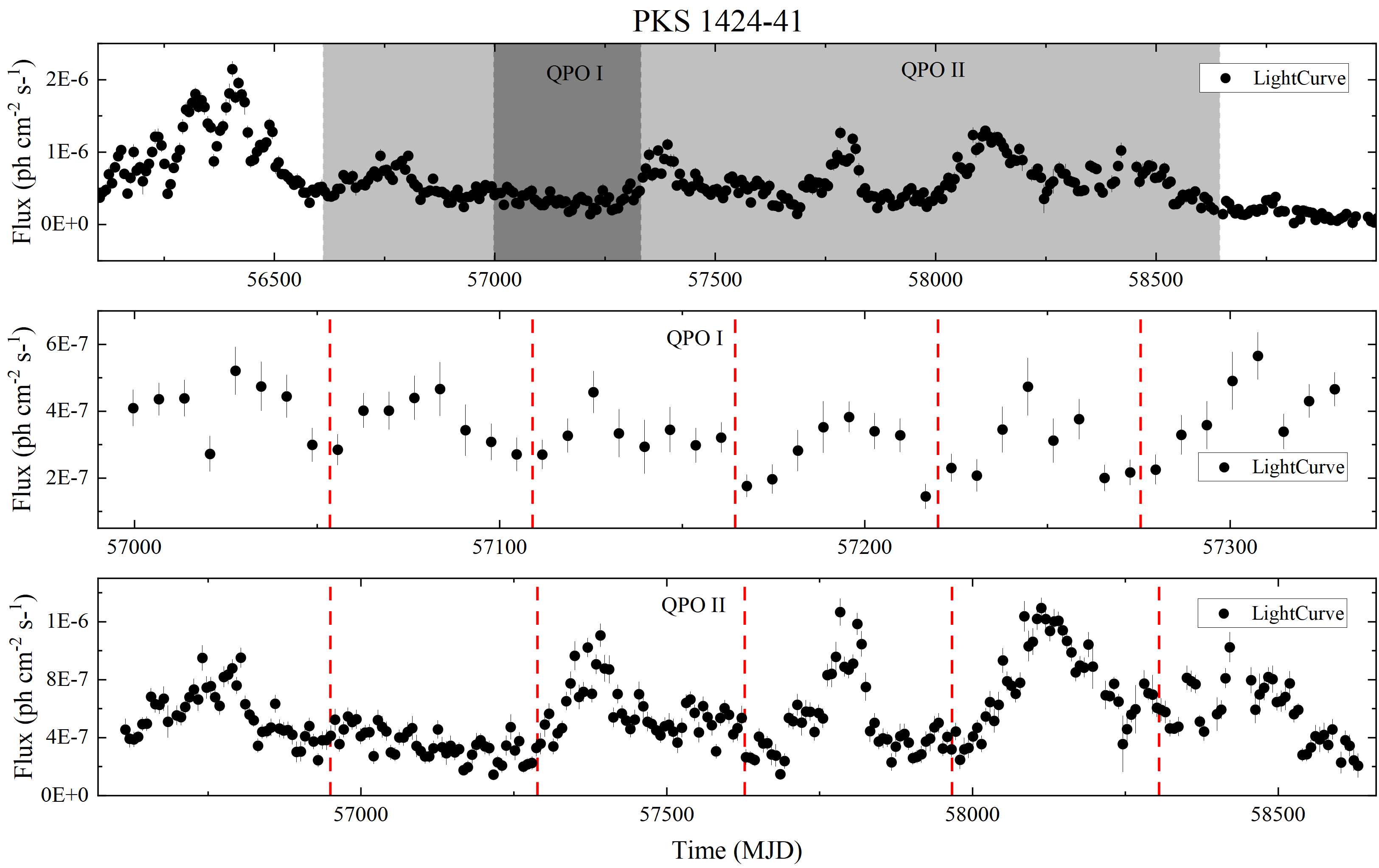}
\end{minipage}
\begin{minipage}[t]{1\textwidth}
\includegraphics[height=6cm,width=7cm]{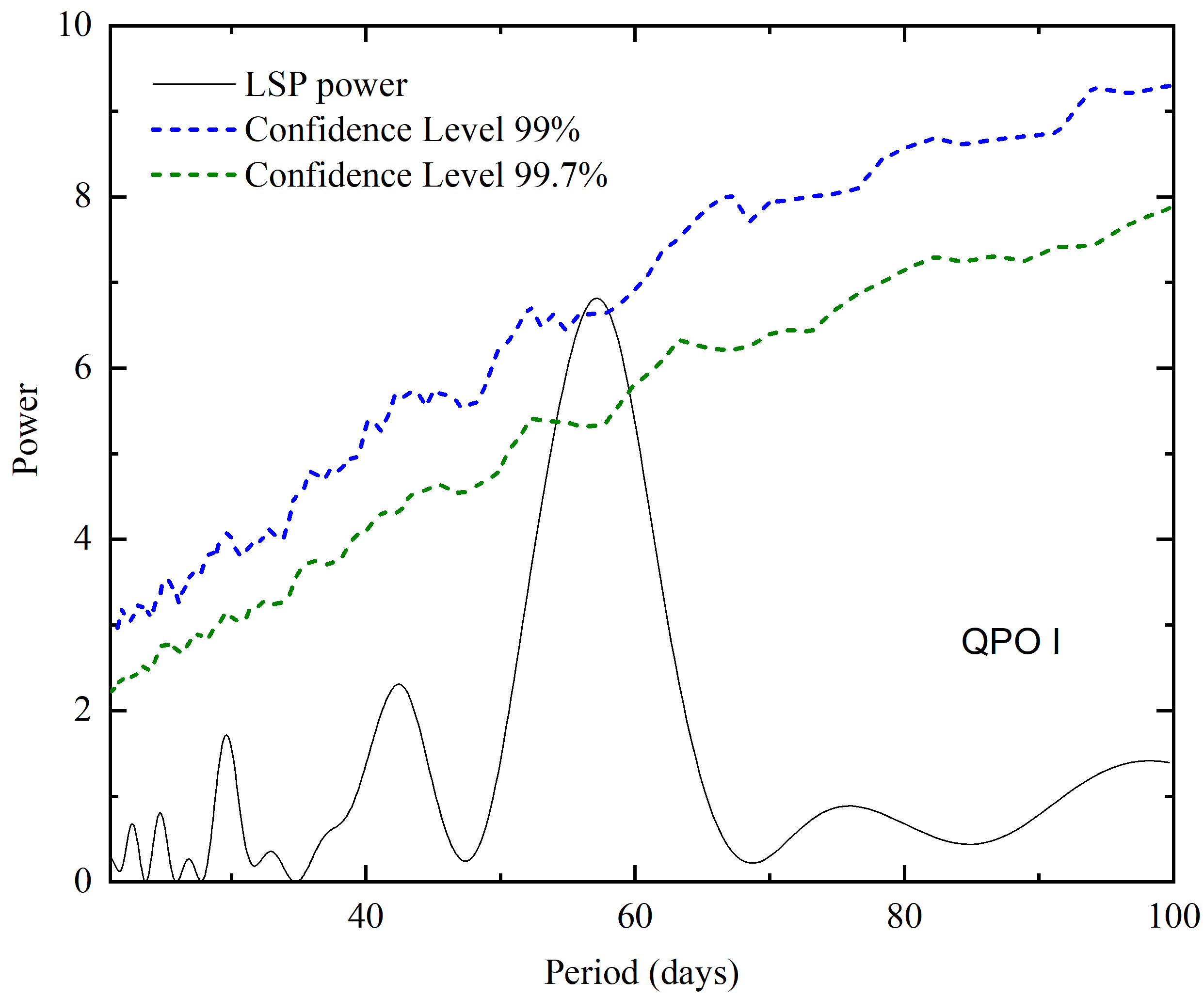}
\includegraphics[height=6cm,width=10.6cm]{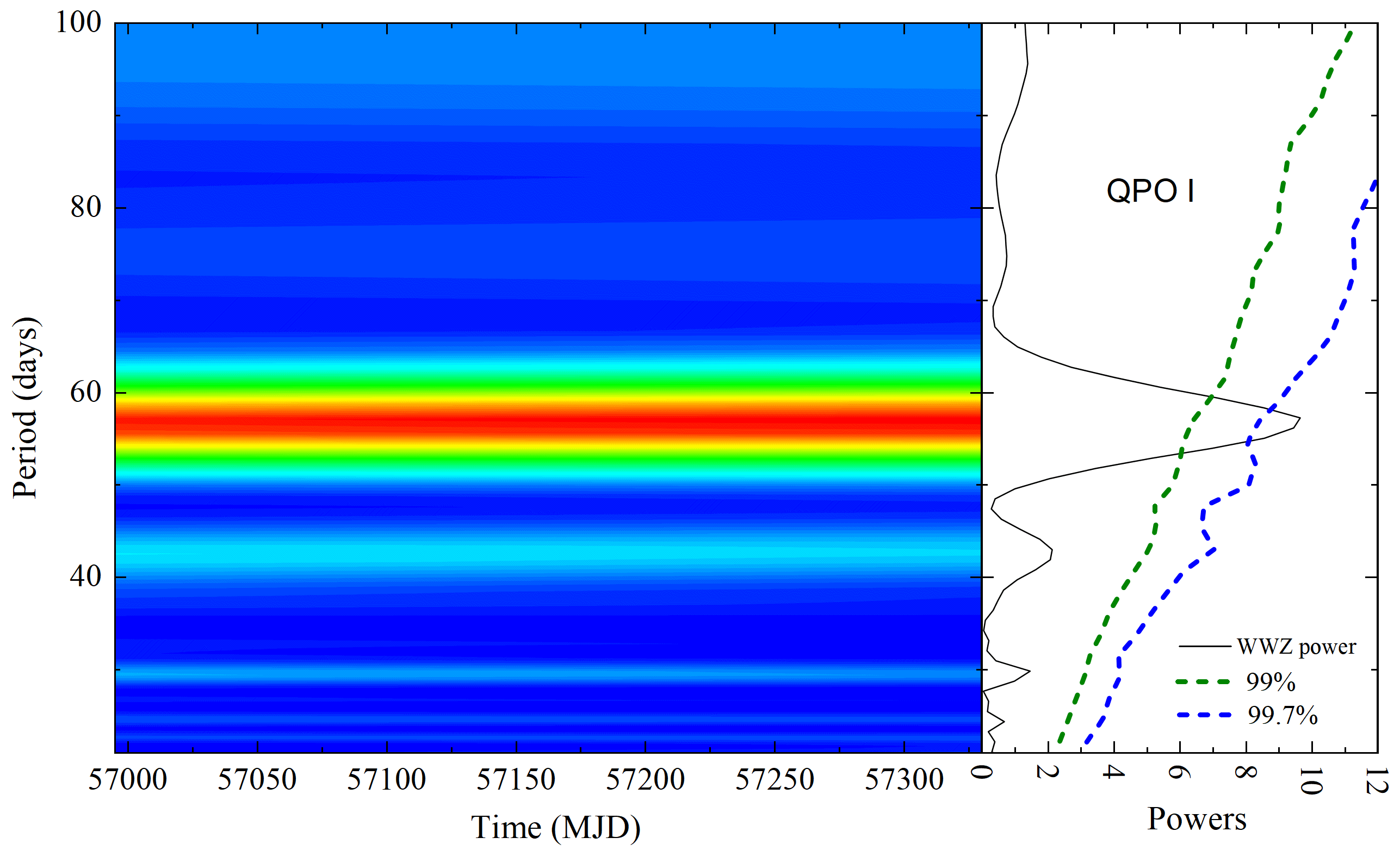}
\end{minipage}
\begin{minipage}[t]{1\textwidth}
\includegraphics[height=6cm,width=7cm]{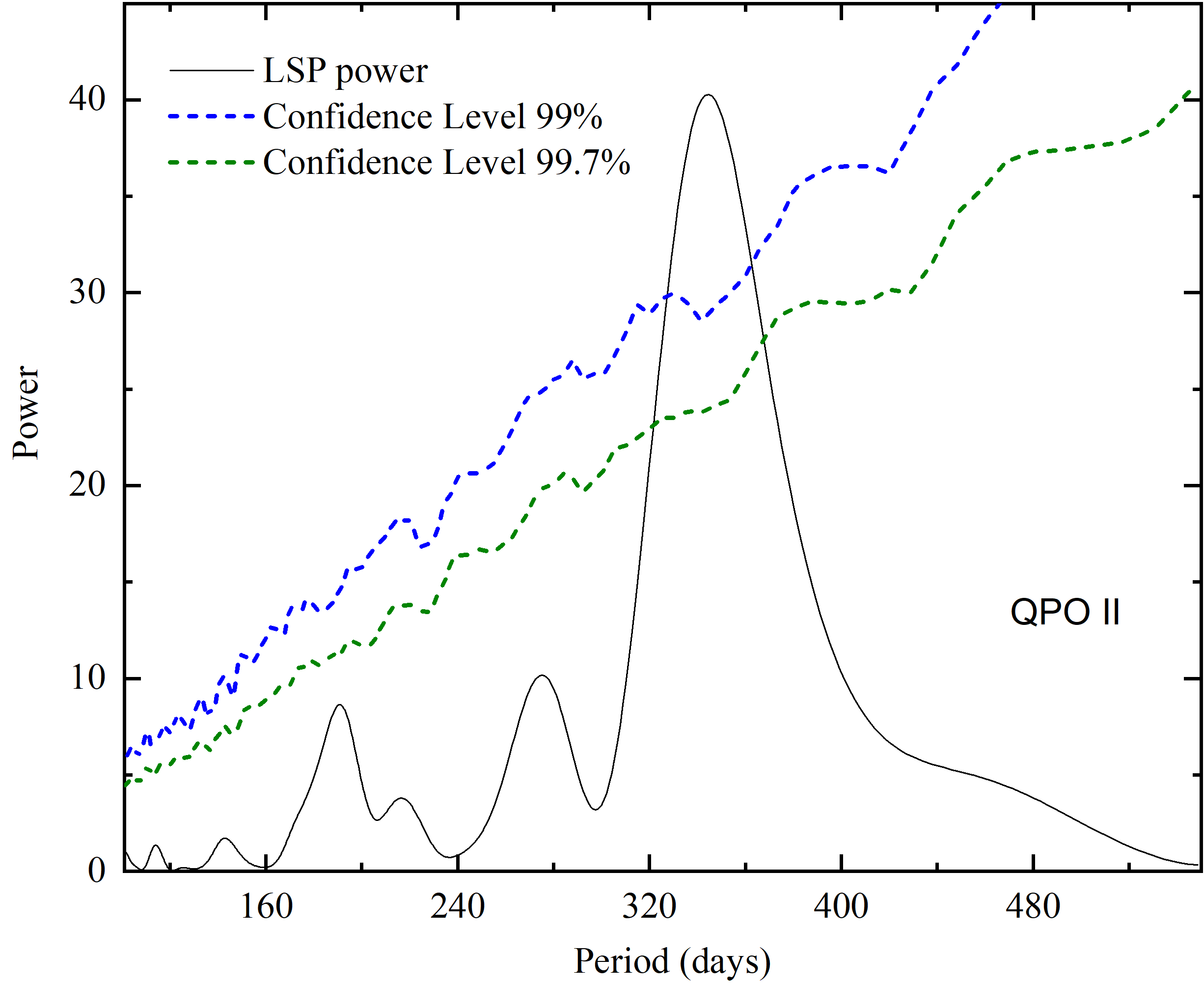}
\includegraphics[height=6.2cm,width=10.6cm]{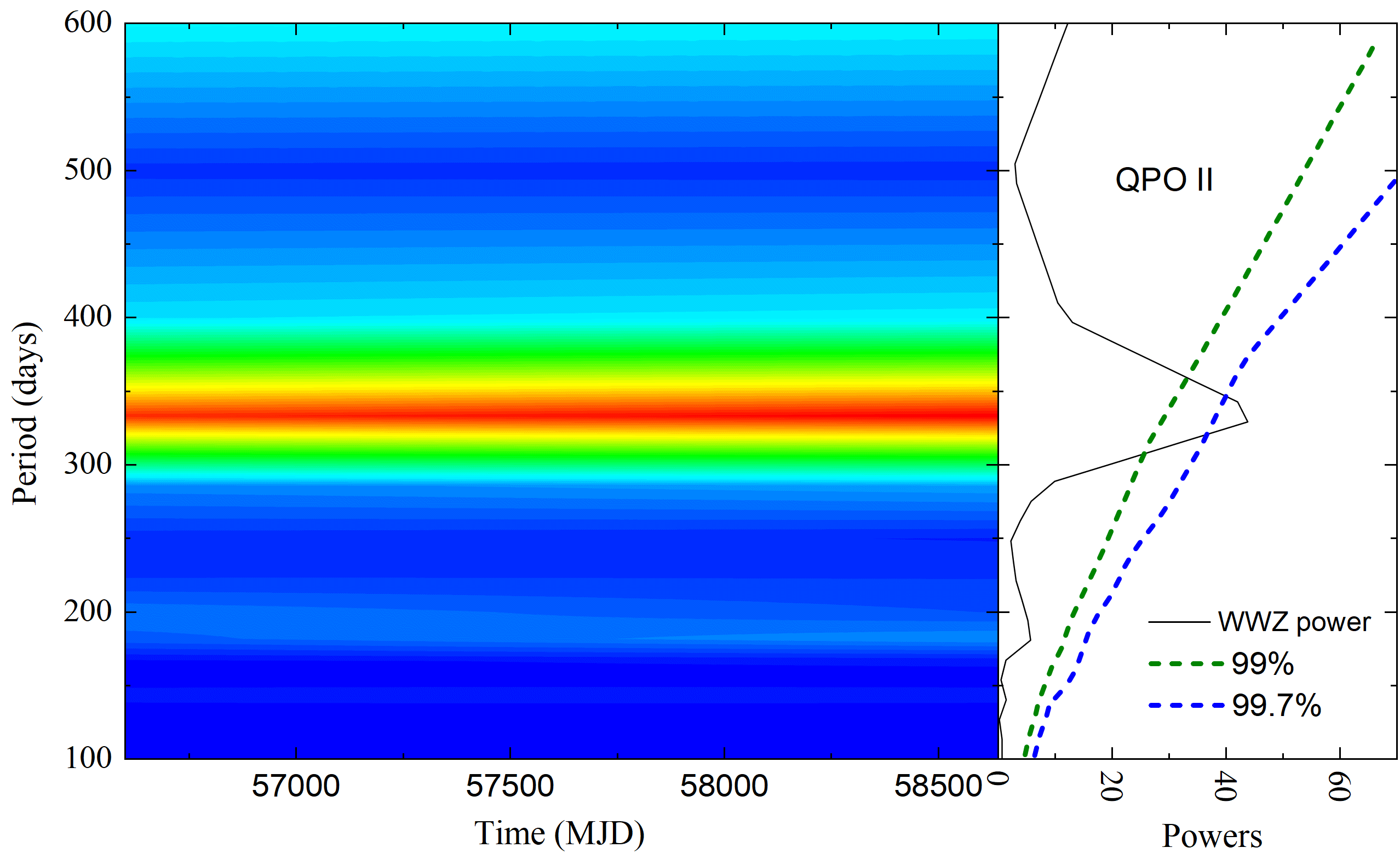}
\end{minipage}
\caption{Same description as Figure \ref{Figure2} for PKS 1424-41 (The blue dotted lines in the bottom panel represent the 99.7\% confidence level contours).}
\label{Figure6}
\end{figure*}

PKS 0537-441 is a BL Lac object with a redshift of 0.89, and its light curve is shown in Figure \ref{Figure5}. Earlier studies showed that the source was active in infrared, optical, ultraviolet, and X-ray, and brightened by about a factor of 2 in all bands \citep{1986ApJ...311L..13T}.
PKS 0537-441 shows a strong flux density variation in the 1.42 GHz band and erupts at J.D.2449011, characterized by a volatility index of about 15\%, a variation of about 45\%, and a time scale of about $10^4$ s. This extremely violent behavior may be the result of strong scattering from dense ionized structures in the interstellar medium \citep{1994A&A...288..731R}.
\cite{2016AJ....151...54S} studied the long-term light curve of PKS 0537-441 and reported the long-term QPO of 285 days for gamma rays for the first time based on the LSP method, and \cite{2023A&A...672A..86R} gave similar results with the CWT method. Their results all prove that there is an annual QPO in the long-term light curve of PKS 0537-441.
Different from the above studies, we are more interested in the transient QPO in the light curve. As shown in Figure \ref{Figure5}, we report for the first time two transient QPOs in PKS 0537-441, between MJD 56803-57183 and MJD 57636-58036, with periods of 55.0 $\pm$ 3.3 and 54.7 $\pm$ 3.3 days, respectively, with seven cycles. These two periods are very close to about 55 days, and there happens to be a relatively obvious Flare state in the middle of them. These two transient QPOs may reveal a very interesting regular radiation process inside the PKS 0537-441.

\subsection{PKS 1424-41}
PKS 1424-418 is an FSRQ with a redshift of 1.524. A study by \cite{2013A&A...555A...2N} showed a clear correlation between NIR and gamma-ray emissions. \cite{2014A&A...569A..40B} analyzed in detail the four flares in the multiband light curve of PKS 1424-418 from 2008 to 2011, showing a high degree of correlation from optical to gamma-ray energy ranges. For all flares, the SED can be adequately represented by a Lepton model that includes inverse Compton emission from an external radiation field with similar parameters.
\cite{2021PASP..133b4101Y} reported a 355-day QPO between MJD 54963-58743 in the gamma-ray band at PKS 1424-418. Later, \cite{2023A&A...672A..86R} also reported a 94-day transient QPO between MJD 56100-56500. The results of our systematic analysis showed that there were 57.2 $\pm$ 4.1 and 341 $\pm$ 25.8 days of transient QPO between MJD 56998-57331 and MJD 57636-58036, respectively, with six cycles. Among them, the 341-day cycle is very close to the 355-day cycle of \cite{2021PASP..133b4101Y}, the 57-day transient QPO is reported for the first time.

\subsection{PKS 1510-089}
\begin{figure*}
\begin{minipage}[t]{1\textwidth}
\centering
\includegraphics[height=7cm,width=14cm]{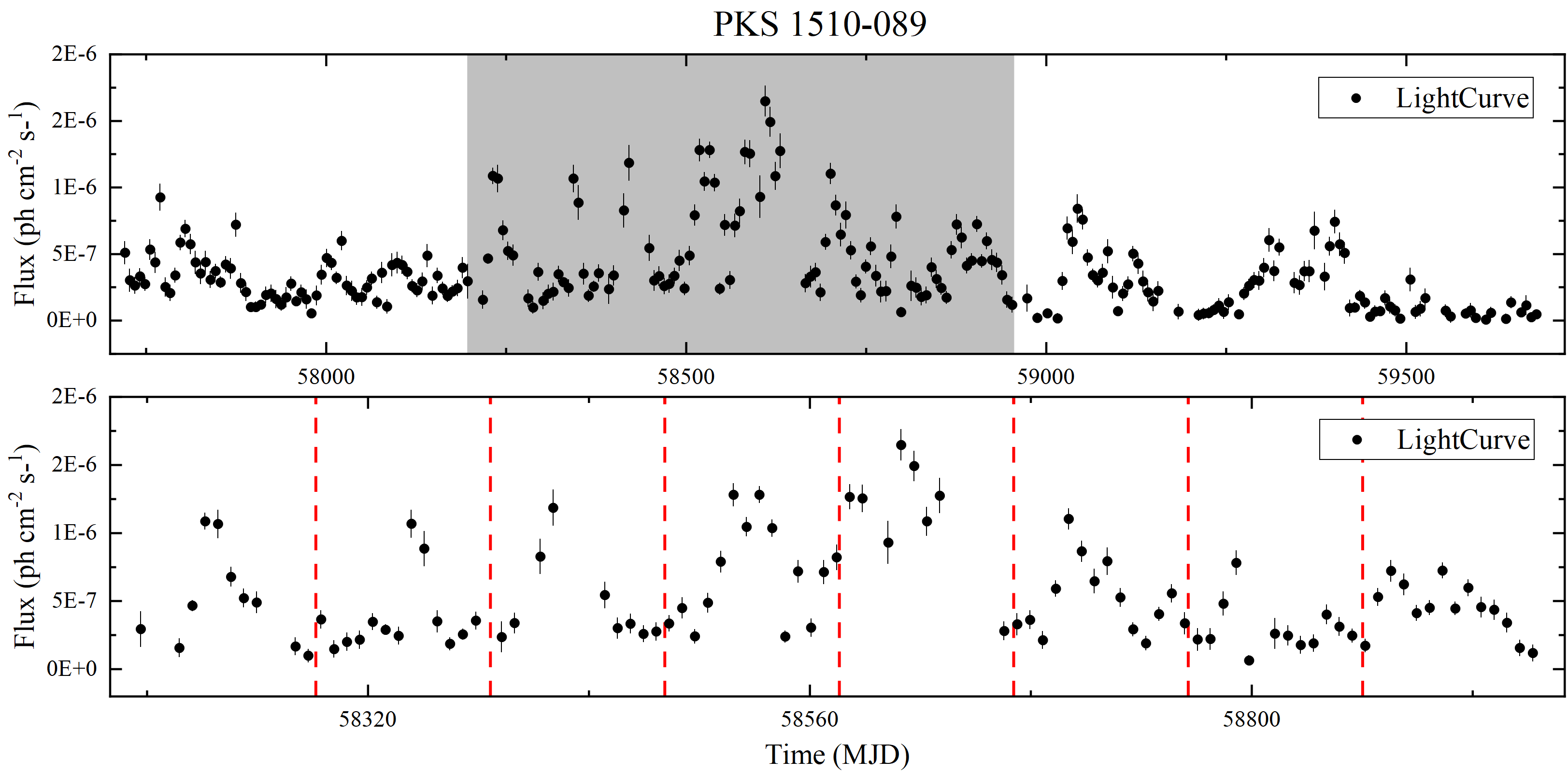}
\end{minipage}
\begin{minipage}[t]{1\textwidth}
\includegraphics[height=6cm,width=7cm]{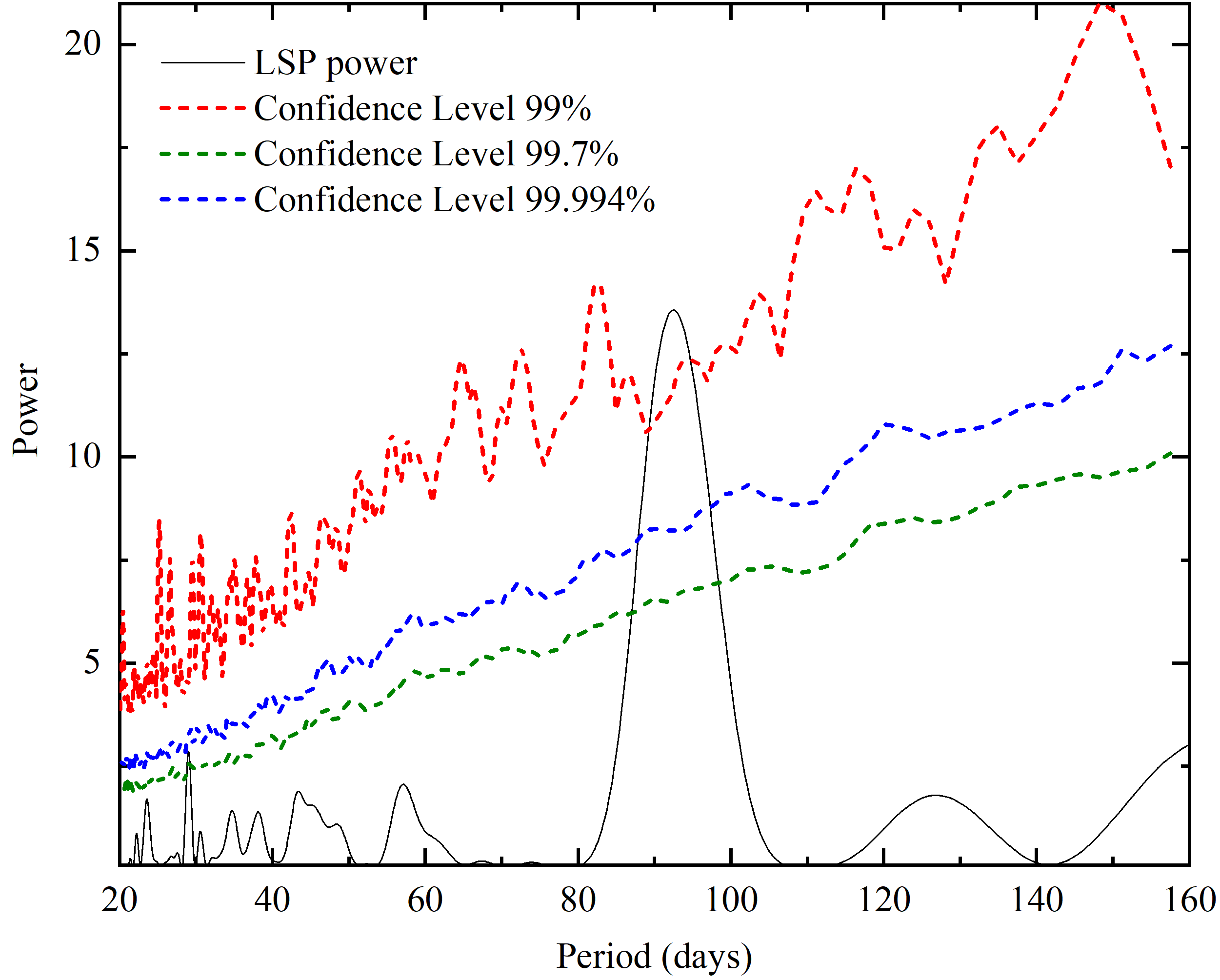}
\includegraphics[height=6.2cm,width=10.6cm]{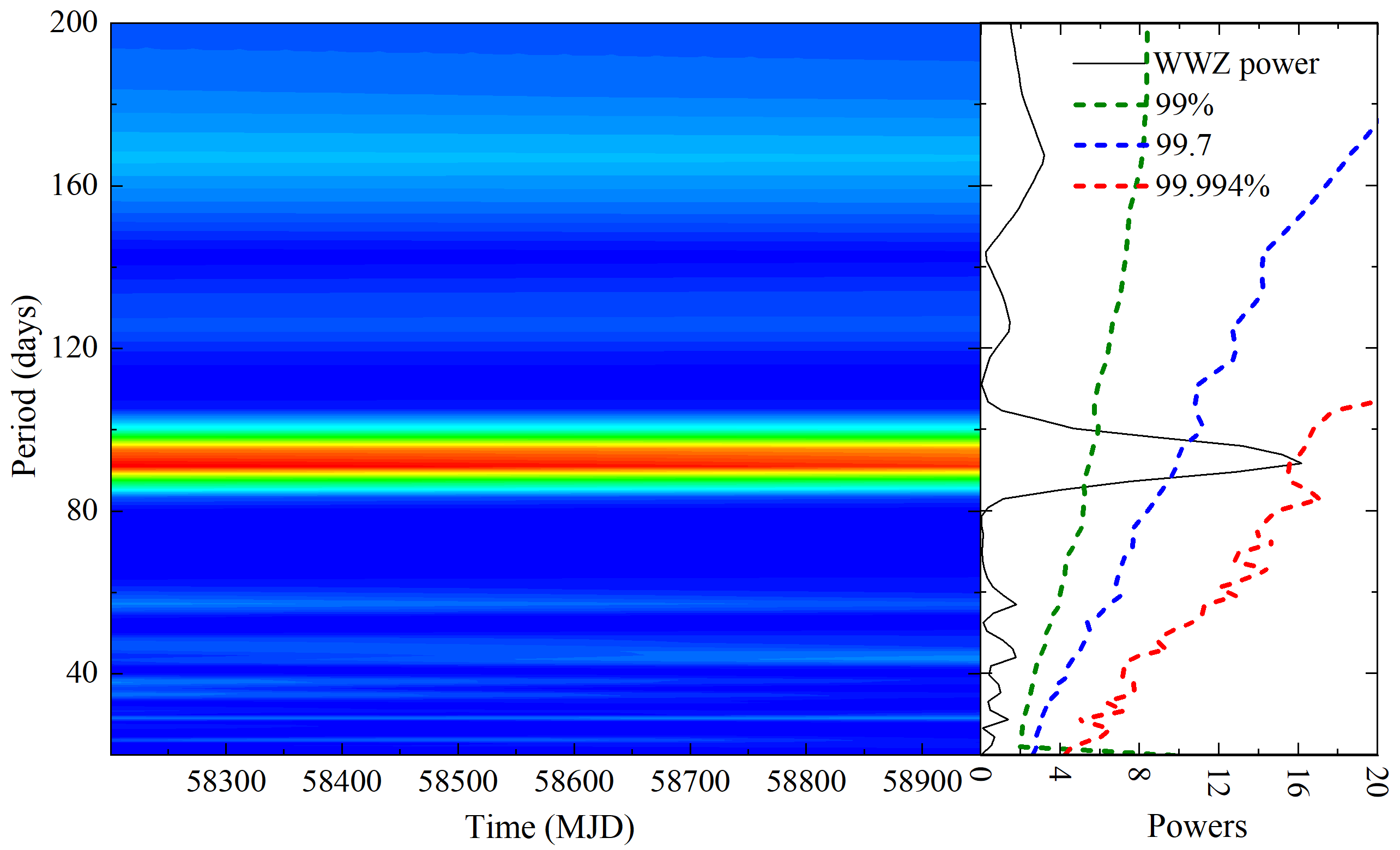}
\end{minipage}
\caption{Same description as Figure \ref{Figure2} for PKS 1510-089 (The blue and red dashed lines in the bottom panel represent the 99.7\% and 99.99\% confidence level contours, respectively).}
\label{Figure7}
\end{figure*}

PKS 1510-089 is a very bright FSRQ with a redshift of 0.36, which has attracted attention because of its high variability, and its multi-wavelength has been detected and widely discussed by many telescopes for a long time. 
In recent years, \cite{2016AJ....151...54S, 2017A&A...601A..30C, 2017A&A...603A..29A, 2021AAS...23752904O, 2022MNRAS.510.3641R, 2023MNRAS.519.4893L} have discussed the QPO of the PKS 1510-089 gamma-ray-radio band. These results are very interesting and have a certain significance for the study of multi-band light variability and the radiation mechanism of this source. In the gamma-ray that this paper focuses on, a QPO of 115 $\pm$ 5 days was reported by \cite{2016AJ....151...54S}. \cite{2022MNRAS.510.3641R} reported QPOs of 3.6 and 92 days in the PKS 1510-089 gamma-ray band of MJD 54906-54923 and MJD 58200-58850, with five and seven cycles, respectively, and they discussed several possible physical models.
The result we give is 91.8 $\pm$ 1.8 days similar to that of \cite{2022MNRAS.510.3641R}. The difference is that the time range can be extended to MJD 58197-58955, with eight cycles, and the confidence level is about 4.3 $\sigma$, as shown in Figure \ref{Figure7}. This result undoubtedly confirms the QPO of about 92-day for the gamma-ray of PKS 1510-089.

\section{Discussion} \label{section:5}

\subsection{Compare with previous works}
The research on the QPO phenomenon in the light curve of blazars has always been a very interesting subject. \cite{2018NatCo...9.4599Z} reported the first case of a transient QPO on a monthly timescale (34.5 days) in blazar PKS 2247-131, which has attracted widespread attention from the astronomical community. Subsequently, transient QPO events on day-to-month time scales have been reported in different sources \citep[see, e.g.,][]{2019MNRAS.484.5785G, 2020A&A...642A.129S, 2021MNRAS.501...50S, 2022MNRAS.510.3641R, 2022ApJ...938....8C}. However, so far, high-confidence transient QPO cases are still rare, and the analysis methods used in different works varies, preventing systematic analysis of the sample. For this reason, an research effort to systematically search for transient QPOs, especially to limit the incidence of transient QPOs, would be of great importance. 

Compared with long-term stable QPOs, searching for transient QPOs as this paper focuses on is more complicated, mainly because the time scale of the cycle for transient QPOs is shorter. The time scale of long-term QPO reported so far is generally more than one year or even more than ten years, while the time scale of transient QPO is generally within one year, mainly tens of days. This leads to higher requirements on the search method. Common period search methods such as power density spectroscopy (PDS), REDFIT, LSP, etc. are difficult to identify the transient QPO in a specific time period in the light curve. Different from these methods, thanks to the flexibility of wavelets, the WWZ method can identify the transient QPO signal in a specific time period in the light curve, but there are still cases of false positives, which generally require careful visual inspection and more detailed analysis to be more convincing. 

In a very recent study, \cite{2023A&A...672A..86R} analyzed the Fermi-LAT light curves of the 35 brightest blazars from the start of the Fermi mission (August 2008) to April 2021, energies from 100 MeV to 300 GeV. Two binning strategies, 7-day and 30-day binning light curves, were used in the analysis. Finally, 24 candidate quasars with transient QPO were identified in 35 active galactic nuclei, and their QPO periods were mainly concentrated within one year. Different from the work of \cite{2023A&A...672A..86R}, here we use a larger search sample in order to search based on a flux-limited complete sample as much as possible, and we use the WWZ method that is more sensitive to transient QPOs. In addition, we use more conservative (strict) criteria to determine the transient QPO: (1) Requiring the number of cycles of the QPO to be greater than or at least equal to five. (2) The data points in the light curve are required to be as complete as possible (those sources whose data points in the light curve are missing more than 60\% are discarded). (3) The periodicity of those light curves that cannot be identified by the naked eye is discarded. Finally, we identified eight transient QPO candidates in six sources, and the confidence level of six of these transient QPO candidates quantitatively assessed by Monte Carlo simulation was not less than 3$\sigma$. The 92-day transient QPO in PKS 1510-089 was consistent with the report in \cite{2022MNRAS.510.3641R}. 
The transient QPOs in 4C +01.02 (253-day), PKS 0402-362 (103-day) and PKS 1424-41 (341-day) are close to the reports in \cite{2021PASP..133b4101Y} and \cite{2023A&A...672A..86R}. Moreover, we found for the first time a 94-day transient QPO in PKS 0336-01, 55-day and 54-day QPOs in PKS 0537-441, and a 57-day QPO in PKS 1424-41.

\subsection{Physical explanation}

The generation mechanism of QPO in active galactic nuclei is still controversial, and some interesting physical models have also been proposed. 
The supermassive binary black hole system achieved great success in explaining the $\sim$ 12-year QPO of OJ 287, and subsequently, this model was often used to explain some stable long-duration QPOs \citep{2008Natur.452..851V,2010MNRAS.402.2087V}.
In addition, the continuous jet precession and Lense-Thirring precession of the accretion disk can also explain the long-duration QPO. There are extensive discussions in the literature such as \cite{2000A&A...360...57R,2004ApJ...615L...5R, 1998ApJ...492L..59S,2018MNRAS.474L..81L,2018ApJ...858...82Y}.
For transient QPOs, it is clear that other physical models need to be considered. One possible mechanism is that hotspots rotating in the innermost stable circular orbit of a supermassive black hole dominate the periodic modulation process in the light curves \citep{1991A&A...246...21Z,2009ApJ...690..216G}.
In this model, the rotation of the hot spot will participate in modulating the seed photon field for external inverse Compton scattering (EC) within the jet, thereby modulating the gamma-ray emission. The transient QPO that we observe lasts only a few days due to the Doppler enhancement effect \citep{2022MNRAS.510.3641R}. 
Second, magnetic reconnection events of nearly equally spaced magnetic islands inside the jet can modulate the QPO of BH systems at different scales \citep{2013RAA....13..705H}. This model requires the formation of uniformly equidistant magnetic islands in a region with an Eddington strength magnetic field close to the black hole to periodically enhance the flux and thus generate the observed transient QPOs. \cite{2018ApJ...854L..26S} modeled the extremely fast variability ($\sim$ 5 minutes) of gamma-ray during the FSRQ CTA 102 outburst based on a magnetic reconnection process. This model can explain some short and fast transient QPOs, but explaining transient QPOs over weeks or even months is obviously more difficult and requires more theoretical studies and numerical simulations to explore.

In addition, a model for the helical motion of the enhanced emission region (or blob) inside the jet has been widely discussed in flare transient QPO studies
\citep{2015ApJ...805...91M,2017MNRAS.465..161S,2018NatCo...9.4599Z}. In the one zone leptonic model, the blazar emission is dominated by the emission region (or blob) in the jet, and the seed photons in the emission region (or blob)  produce high-energy gamma-ray emission through a synchrotron self-Compton process. The large-scale magnetic field or some dynamical reasons may cause the jet or the emission region (or blob) within the jet to produce a helical motion, which is also known as the helical jet structure \citep{2015ApJ...805...91M,2017MNRAS.465..161S}.
In this scenario, the viewing angle ($\theta$) changes with the motion of the emission region (or blob), which can be described as
\begin{equation}
cos\theta=cos\phi cos\psi+sin\phi sin\psi cos(2\pi t/P_{obs}),
\label{eq:LebsequeIp7}
\end{equation}
where $\psi$ is the angle between the jet and the observer's line of sight, $\phi$ is the pitch angle of the helical path of the emission region (or blob) with respect to the jet axis, and $P_{obs}$ is the periodicity in the observed light curve \citep{2017MNRAS.465..161S,2018NatCo...9.4599Z}.
For the observer, the gamma-ray emission is enhanced by relativistic clustering effects, so that there is a translation between the true physical driving period $P$ of the emission region (or blob) and the observed period, $P=P_{obs}/(1-\beta cos\phi cos\psi)$, where $\phi$ and $\psi$ we take typical values of $2^{\circ}$ and $5^{\circ}$ \citep{2018NatCo...9.4599Z}.
The bulk Lorentz factor ($\Gamma = 1/(1-\beta^{2})^{1/2}$) of the emission region (or blob), where Lorentz factor we also take some typical values, for FSRQ type and BL lac type we take $\Gamma = 20$ and $\Gamma = 8.5$ respectively. 
Then, by substituting specific parameters on this basis, the real physical driving period $P$ of each transient QPO, the distance $D$ that the emitting region (or sphere) moves in one cycle ($D=c\beta P cos\phi$), and the total projected distance of the entire transient QPO phase $D_p=N Dsin\psi$ can be estimated (see Table \ref{tab:tab3}). In terms of timescales, the helical jet structural scenario provides a more plausible explanation for most of the transient QPOs we discuss here.

\begin{table}
	\centering
	\caption{Helical Jet Parameters for Mentioned Features.}
	\label{tab:tab3}
	\begin{tabular}{lcccr} 
		\hline
		Source name    & $P_{obs}$ (days)  & P (years)     & D (pc)     & $D_p$ (pc) \\
        \hline
        4C +01.02      &  253.2 $\pm$ 20.0    &   $\sim$ 122.6   &   $\sim$ 37.5  &  $\sim$ 19.6   \\
        PKS 0336-01    &  94.6 $\pm$ 6.8    &   $\sim$ 19.6   &   $\sim$ 14.0   &   $\sim$ 7.3 \\
        PKS 0402-362   &  103.5 $\pm$ 7.9    &   $\sim$ 50.1  &   $\sim$ 15.3  &   $\sim$ 6.7 \\
        PKS 0537-441   &  55.0 $\pm$ 3.3  &   $\sim$ 13.3   &   $\sim$ 4.0    &   $\sim$ 2.5 \\
                       &  54.7 $\pm$ 3.3  &   $\sim$ 13.2   &   $\sim$ 4.0    &   $\sim$ 2.5 \\
        PKS 1424-41    &  57.2 $\pm$ 4.1  &   $\sim$ 27.7   &   $\sim$ 8.5    &   $\sim$ 4.4 \\
                       &  341 $\pm$ 25.8  &   $\sim$ 165.1   &   $\sim$ 50.5    &   $\sim$ 26.4 \\
        PKS 1510-089   &  92.3 $\pm$ 5.2    &   $\sim$ 44.7  &   $\sim$ 13.7    &  $\sim$ 9.5 \\
		\hline
	\end{tabular}
\end{table}

In this work, we also note two special cases, PKS 0537-441 and PKS 1424-41. We find a transient QPO of about 55 days at MJD 56803-57183 of PKS 0537-441, and the same transient QPO appears again within the error range about 450 days later at MJD 57636-58036. The reappearance of the transient QPO in PKS 0537-441 in the helical jet scenario implies that the spiraling emitting region (or blob) within the jet is reproducible, in other words, after 450 days the helicalling emitting region (or blob) again modulates the PKS 0537-441 gamma ray emission in the same way. This is the first case of a repeated transient QPO reported to date.

In addition, the two cases of transient QPO in PKS 1424-41 are also very interesting. Transient QPOs of approximately 57 days and approximately 341 days occurred at MJD 56998-57331 and MJD 56611-58644, respectively, with the 57-day transient QPO occurring between the second and third cycles of the 341-day transient QPO. And the occurrence of the 57-day transient QPO causes the peak flux in the second period of the 341-day transient QPO to be lower than the average of the peak flux in the other five periods. Obviously, it is difficult to explain this special nested transient QPO appearing in PKS 1424-41 in a simple helical jet model. If the origin of the 341-day transient QPO is indeed the helical motion of the emission region (or blob) within the jet, then it is likely that the origin of the 57-day transient QPO involves a more intricate set of physical mechanisms.

\subsection{Incidence rate of Transient QPO Events}
Based on the light variability properties of transient QPOs reported here and some transient QPOs reported in the literature, we noticed some interesting and important clues: (1) transient QPOs tend to appear in the flare states of the blazars. (2) Most of the reported transient QPOs in the literature appear to happen mainly in FSRQs, with comparatively fewer occurrences in BL Lacs. Our sample shows a similar pattern. However, whether this phenomenon is due to sample selection bias because of FSRQ's brighter luminosity than BL Lacs in the Fermi-LAT gamma-ray band requires further investigation in the future. (3) Of the 134 bright sources examined in this paper, only 6 were found to have transient QPOs, and of these 6, only 4 exhibited transient QPOs with a confidence level greater than 3$\sigma$. The incidence of high-confidence transient QPO events is about 3\%, which suggests that transient QPO phenomena are very rare in bright flares.

It seems plausible that the appearance of QPO indicates the formation of some ordered low-entropy dynamic structure within the black hole system. Physically, the formation of such an ordered low-entropy structure would require certain specific physical processes within the system, such as the helical motion of the enhanced emission region (or blob) inside the jet, hotspots rotating on the innermost stable circular orbit of the supermassive black hole, or magnetic reconnection between nearly equidistant magnetic islands within the jet, and the probability of its formation depends on these specific physical mechanisms. In the framework of interpreting QPOs as arising from the helical motion of an enhanced emission region (or blob) within a jet, a large-scale stabilizing magnetic field would be required to underlie the jet's helical structure. And the probability of establishing such an ordered large-scale magnetic field configuration would determine the occurrence of transient QPO phenomena. 
Thus, the low rate of transient QPO events at least strongly implies that transient QPO may not be dominated by a universal physical process. Tighter empirical constraints on the occurrence of transient QPO events through future observations, alongside comparative analyses of the incidence of specific processes invoked by different physical models producing transient QPO phenomena, have the potential to provide strong discriminatory power between different theoretical frameworks. 

\section{summary} \label{section:6}

In this paper, a systematic search for transient QPO signals was conducted in 134 blazars with a peak flux exceeding $1\times10^{-6}$~ph~cm$^{-2}$~s$^{-1}$, based on the 0.1-300 GeV gamma-ray light curve from Fermi-LAT. The main results are summarized in the following:

1. This work reports four new transient QPO events, including an approximately 94-day event in PKS 0336-01, two events lasting approximately 55 days each in PKS 0537-441, and an approximately 57-day event in PKS 1424-41.

2. We discussed the physical origin of transient QPOs, and the presence of a helical jet structure in the enhanced emission region provides the best explanation.

3. Unique transient QPO events were identified in two sources. The occurrence of repeated transient QPOs in PKS 0537-441 suggests that the emission region (or blob) with helical motion modulates the gamma-ray emission again. The special nested transient QPO events in PKS 1424-41 may be related to more complex physical scenarios.

4. We found that transient QPO events tend to occur in the flaring state of blazars, and they are more likely to occur in the population of flat-spectrum radio quasars. Additionally, the occurrence rate of transient QPO events is estimated to be approximately 3\%.

\section*{DATA AVAILABILITY}
\noindent Monitored Source List Light Curves: \url{https://fermi.gsfc.nasa.gov/ssc/data/access/lat/msl_lc/} \\
\noindent The Fermi-LAT data used in this article are available in the LAT data server at: \url{https://fermi.gsfc.nasa.gov/ssc/data/access/} \\
\noindent The Fermi-LAT data analysis software is available at: \url{https://fermi.gsfc.nasa.gov/ssc/data/analysis/software/} \\
\noindent The weighted wavelet Z-transform method: \url{https://github.com/eaydin/WWZ/} \\
\noindent This light curve simulation method: \url{https://github.com/samconnolly/DELightcurveSimulation/},

\section*{Acknowledgements}
We are grateful for the publicly available data from Fermi-LAT and the standard analysis procedures provided by the LAT collaboration. The work is supported by the National Natural Science Foundation of China (No. 12103022) and the Special Basic Cooperative Research Programs of Yunnan Provincial Undergraduate Universities’ Association (No. 202101BA070001-043). 
Nan Ding is sincerely grateful for the financial support of the Xingdian Talents Support Program, Yunnan Province (NO. XDYC-QNRC-2022-0613). 


\appendix
\section{Fermi-LAT bright source samples}
\onecolumn
\begin{longtable}{lccccc}
\caption{Fermi-LAT bright source samples.}  \label{tab:tab4}\\
\hline
Source name  &  4FGL name  &  Source class  &  R.A.(J2000)  &  Dec.(J2000)  &  Redshift \\
\hline
\endfirsthead
\multicolumn{6}{c}%
{{ \tablename\ \thetable{} -- continued Fermi-LAT bright source samples.}} \\
\hline
Source name  &  4FGL name  &  Source class  &  R.A.(J2000)  &  Dec.(J2000)  &  Redshift \\
\hline
\endhead
\hline \multicolumn{6}{c}{{Continued on next page}} \\ 
\endfoot
\endlastfoot
PMN J0017-0512   &   4FGL J0017.5-0514   &   FSRQ   &   4.3992   &   -5.2116   &   0.227\\
TXS 0059+581   &   4FGL J0102.8+5824   &   FSRQ   &   15.6907   &   58.4031   &   0.644\\
4C +01.02   &   4FGL J0108.6+0134   &   FSRQ   &   17.1615   &   1.5834   &   2.099\\
4C 31.03   &   4FGL J0112.8+3208   &   FSRQ   &   18.2097   &   32.1382   &   0.603\\
PKS 0130-17   &   4FGL J0132.7-1654   &   FSRQ   &   23.1812   &   -16.9135   &   1.02\\
PKS 0131-522   &   4FGL J0133.1-5201   &   FSRQ   &   23.274   &   -52.0011   &   0.02\\
PKS 0202-17   &   4FGL J0205.0-1700   &   FSRQ   &   31.2403   &   -17.0222   &   1.7395\\
0208-512   &   4FGL J0210.7-5101   &   FSRQ   &   32.6925   &   -51.0172   &   0.999\\
CGRaBS J0211+1051   &   4FGL J0211.2+1051   &   BL Lac   &   32.8049   &   10.8597   &   0.2\\
S3 0218+35   &   4FGL J0221.1+3556   &   FSRQ   &   35.2729   &   35.9372   &   0.944\\
3C 66A   &   4FGL J0222.6+4302   &   BL Lac   &   35.665   &   43.0355   &   0.37\\
PKS 0226-559   &   4FGL J0228.3-5547   &   FSRQ   &   37.09   &   -55.7676   &   2.464\\
4C +28.07   &   4FGL J0237.8+2848   &   FSRQ   &   39.4684   &   28.8025   &   1.213\\
0235+164   &   4FGL J0238.6+1637   &   BL Lac   &   39.6622   &   16.6165   &   0.94\\
PKS 0244-470   &   4FGL J0245.9-4650   &   FSRQ   &   41.5005   &   -46.8548   &   1.385\\
PKS 0250-225   &   4FGL J0252.8-2219   &   FSRQ   &   43.1998   &   -22.3237   &   1.427\\
PKS 0301-243   &   4FGL J0303.4-2407   &   BL Lac   &   45.8604   &   -24.1198   &   0.266\\
PKS 0336-01   &   4FGL J0339.5-0146   &   FSRQ   &   54.8789   &   -1.7766   &   0.852\\
PKS 0346-27   &   4FGL J0348.5-2749   &   FSRQ   &   57.1589   &   -27.8204   &   0.991\\
4C +50.11   &   4FGL J0359.6+5057   &   FSRQ   &   59.916   &   50.959   &   1.52\\
PKS 0402-362   &   4FGL J0403.9-3605   &   FSRQ   &   60.974   &   -36.0839   &   1.4228\\
PKS 0426-380   &   4FGL J0428.6-3756   &   BL Lac   &   67.1684   &   -37.9388   &   1.11\\
NRAO 190   &   4FGL J0442.6-0017   &   FSRQ   &   70.6611   &   -0.2954   &   0.844\\
PKS 0454-234   &   4FGL J0457.0-2324   &   FSRQ   &   74.2632   &   -23.4145   &   1.003\\
PKS 0458-02   &   4FGL J0501.2-0158   &   FSRQ   &   75.3034   &   -1.9873   &   2.286\\
PKS 0502+049   &   4FGL J0505.3+0459   &   FSRQ   &   76.3466   &   4.9952   &   0.954\\
PKS 0514-459   &   4FGL J0515.6-4556   &   FSRQ   &   78.9385   &   -45.9453   &   0.194\\
VER 0521+211   &   4FGL J0521.7+2112   &   BL Lac   &   80.445   &   21.213   &   0.108\\
CRATES J0531-4827   &   4FGL J0532.0-4827   &   BL Lac   &   83.002   &   -48.461   &   0.8116\\
OG 050   &   4FGL J0532.6+0732   &   FSRQ   &   83.1625   &   7.5454   &   1.254\\
PKS 0537-441   &   4FGL J0538.8-4405   &   BL Lac   &   84.7098   &   -44.0858   &   0.89\\
PKS 0537-286   &   4FGL J0539.9-2839   &   FSRQ   &   84.9762   &   -28.6655   &   3.104\\
TXS 0552+398   &   4FGL J0555.6+3947   &   FSRQ   &   88.901   &   39.788   &   2.365\\
TXS 0646-176   &   4FGL J0648.4-1743   &   FSRQ   &   102.1187   &   -17.7348   &   1.232\\
MG2 J071354+1934   &   4FGL J0713.8+1935   &   FSRQ   &   108.482   &   19.5834   &   0.54\\
S5 0716+714   &   4FGL J0721.9+7120   &   BL Lac   &   110.488   &   71.341   &   0.31\\
4C 14.23   &   4FGL J0725.2+1425   &   FSRQ   &   111.32   &   14.4205   &   1.039\\
PKS 0727-11   &   4FGL J0730.3-1141   &   FSRQ   &   112.5796   &   -11.6868   &   1.591\\
PKS 0736+01   &   4FGL J0739.2+0137   &   FSRQ   &   114.8251   &   1.6179   &   0.1894\\
BZU J0742+5444   &   4FGL J0742.6+5443   &   FSRQ   &   115.671   &   54.727   &   0.72\\
B2 0748+33   &   4FGL J0752.2+3313   &   FSRQ   &   118.055   &   33.232   &   1.9352\\
PKS 0805-07   &   4FGL J0808.2-0751   &   FSRQ   &   122.0647   &   -7.8527   &   1.837\\
TXS 0827+243   &   4FGL J0830.8+2410   &   FSRQ   &   127.701   &   24.173   &   0.9388\\
S5 0836+71   &   4FGL J0841.3+7053   &   FSRQ   &   130.3515   &   70.895   &   2.172\\
PMN J0852-5755   &   4FGL J0852.5-5755   &   BCU   &   133.1614   &   -57.9249   &   ... \\
OJ 287   &   4FGL J0854.8+2006   &   BL Lac   &   133.7036   &   20.1085   &   0.3056\\
3EG J0903-3531   &   4FGL J0904.5-3513   &   BCU   &   136.14   &   -35.231   &   ... \\
PKS 0903-57   &   4FGL J0904.9-5734   &   BCU   &   136.2216   &   -57.5849   &   0.695\\
PKS B0906+015   &   4FGL J0909.1+0121   &   FSRQ   &   137.292   &   1.3599   &   1.0232\\
S5 0917+62   &   4FGL J0921.6+6216   &   FSRQ   &   140.418   &   62.271   &   1.4576\\
PKS 0920-39   &   4FGL J0922.7-3959   &   FSRQ   &   140.692   &   -39.987   &   0.591\\
S4 0954+65   &   4FGL J0958.7+6534   &   BL Lac   &   149.6969   &   65.5652   &   0.368\\
PKS 1004-217   &   4FGL J1006.7-2159   &   FSRQ   &   151.6934   &   -21.989   &   0.331\\
4C +40.25   &   4FGL J1023.1+3949   &   FSRQ   &   155.7982   &   39.8043   &   1.2545\\
S4 1030+61   &   4FGL J1033.9+6050   &   FSRQ   &   158.4643   &   60.852   &   1.4082\\
PMN J1038-5311   &   4FGL J1038.8-5312   &   FSRQ   &   159.6694   &   -53.1954   &   1.45\\
S5 1044+71   &   4FGL J1048.4+7143   &   FSRQ   &   162.1151   &   71.7266   &   1.15\\
4C +01.28   &   4FGL J1058.4+0133   &   BL Lac   &   164.6234   &   1.5663   &   0.8936\\
Mrk 421   &   4FGL J1104.4+3812   &   BL Lac   &   166.1138   &   38.2088   &   0.0301\\
PMN J1123-6417   &   4FGL J1123.5-6418   &   BCU   &   170.889   &   -64.311   &   ... \\
PKS 1127-14   &   4FGL J1129.8-1447   &   FSRQ   &   172.5294   &   -14.8243   &   1.184\\
B2 1144+40   &   4FGL J1146.9+3958   &   FSRQ   &   176.74   &   39.977   &   1.088\\
Ton 599   &   4FGL J1159.5+2914   &   FSRQ   &   179.8826   &   29.2455   &   0.7247\\
1ES 1215+303   &   4FGL J1217.9+3007   &   BL Lac   &   184.467   &   30.1168   &   0.129\\
TXS 1219+285   &   4FGL J1221.5+2814   &   BL Lac   &   185.382   &   28.2329   &   0.102\\
PKS B1222+216   &   4FGL J1224.9+2122   &   FSRQ   &   186.2269   &   21.3796   &   0.4338\\
3C 273   &   4FGL J1229.0+0202   &   FSRQ   &   187.2779   &   2.0524   &   0.1583\\
ON 246   &   4FGL J1230.2+2517   &   BL Lac   &   187.5587   &   25.302   &   0.14\\
J123939+044409   &   4FGL J1239.5+0443   &   FSRQ   &   189.885   &   4.728   &   1.7606\\
PKS 1244-255   &   4FGL J1246.7-2548   &   FSRQ   &   191.695   &   -25.797   &   0.633\\
3C 279   &   4FGL J1256.1-0547   &   FSRQ   &   194.0465   &   -5.7893   &   0.5362\\
OP 313   &   4FGL J1310.5+3221   &   FSRQ   &   197.6194   &   32.3455   &   0.9959\\
GB6 B1310+4844   &   4FGL J1312.6+4828   &   BCU   &   198.1806   &   48.4753   &   0.501\\
PKS 1313-333   &   4FGL J1316.1-3338   &   FSRQ   &   199.0333   &   -33.6498   &   1.21\\
PKS 1335-127   &   4FGL J1337.6-1257   &   FSRQ   &   204.4158   &   -12.9569   &   0.539\\
B3 1343+451   &   4FGL J1345.5+4453   &   FSRQ   &   206.3882   &   44.8832   &   2.542\\
1406-076   &   4FGL J1408.9-0751   &   FSRQ   &   212.2353   &   -7.8741   &   1.5\\
OQ 334   &   4FGL J1422.3+3223   &   FSRQ   &   215.6266   &   32.3862   &   0.6816\\
PKS 1424-41   &   4FGL J1427.9-4206   &   FSRQ   &   216.9846   &   -42.1054   &   1.522\\
H 1426+428   &   4FGL J1428.5+4240   &   BL Lac   &   217.1359   &   42.6725   &   0.129\\
PKS 1454-354   &   4FGL J1457.4-3539   &   FSRQ   &   224.3613   &   -35.6528   &   1.424\\
PKS 1502+106   &   4FGL J1504.4+1029   &   FSRQ   &   226.1041   &   10.4942   &   1.838\\
B2 1504+37   &   4FGL J1506.1+3731   &   FSRQ   &   226.5397   &   37.5142   &   0.6715\\
PKS 1510-089   &   4FGL J1512.8-0906   &   FSRQ   &   228.2106   &   -9.1   &   0.36\\
PKS 1514-24   &   4FGL J1517.7-2422   &   BL Lac   &   229.4242   &   -24.3721   &   0.049\\
B2 1520+31   &   4FGL J1522.1+3144   &   FSRQ   &   230.5416   &   31.7373   &   1.489\\
PG 1553+113   &   4FGL J1555.7+1111   &   BL Lac   &   238.9294   &   11.1901   &   0.36\\
PKS B 1622-297   &   4FGL J1626.0-2950   &   FSRQ   &   246.5251   &   -29.8575   &   0.815\\
1633+382   &   4FGL J1635.2+3808   &   FSRQ   &   248.817   &   38.14   &   1.814\\
Mrk 501   &   4FGL J1653.8+3945   &   BL Lac   &   253.4676   &   39.7602   &   0.0329\\
GB6 J1700+6830   &   4FGL J1700.0+6830   &   FSRQ   &   255.022   &   68.504   &   0.301\\
B3 1708+433   &   4FGL J1709.7+4318   &   FSRQ   &   257.4212   &   43.3124   &   1.027\\
PMN J1717-5155   &   4FGL J1717.6-5154   &   FSRQ   &   259.403   &   -51.909   &   ... \\
S4 1726+45   &   4FGL J1727.4+4530   &   FSRQ   &   261.8652   &   45.511   &   0.717\\
1730-130   &   4FGL J1733.0-1305   &   FSRQ   &   263.263   &   -13.086   &   0.902\\
OT 355   &   4FGL J1734.3+3858   &   FSRQ   &   263.5857   &   38.9643   &   0.97\\
S4 1749+70   &   4FGL J1748.6+7005   &   BL Lac   &   267.1368   &   70.0974   &   0.77\\
OT 081   &   4FGL J1751.5+0938   &   BL Lac   &   267.8867   &   9.6502   &   0.322\\
S5 1803+78   &   4FGL J1800.6+7828   &   BL Lac   &   270.1903   &   78.4678   &   0.68\\
S4 1800+44   &   4FGL J1801.5+4404   &   FSRQ   &   270.3846   &   44.0728   &   0.663\\
PKS 1824-582   &   4FGL J1829.2-5813   &   FSRQ   &   277.3017   &   -58.232   &   1.531\\
PKS 1830-211   &   4FGL J1833.6-2103   &   FSRQ   &   278.416   &   -21.0615   &   2.507\\
CGRaBS J1849+6705   &   4FGL J1849.2+6705   &   FSRQ   &   282.317   &   67.0949   &   0.657\\
PKS B1908-201   &   4FGL J1911.2-2006   &   FSRQ   &   287.7902   &   -20.1153   &   1.119\\
TXS 1923+123   &   4FGL J1925.7+1227   &   BCU   &   291.4201   &   12.4606   &   ... \\
PKS 1954-388   &   4FGL J1958.0-3845   &   FSRQ   &   299.4992   &   -38.7518   &   0.63\\
1ES 1959+650   &   4FGL J2000.0+6508   &   BL Lac   &   299.9994   &   65.1485   &   0.047\\
S5 2007+77   &   4FGL J2005.5+7752   &   BL Lac   &   301.3792   &   77.8787   &   0.342\\
PKS 2023-07   &   4FGL J2025.6-0735   &   FSRQ   &   306.4194   &   -7.598   &   1.388\\
PKS 2032+107   &   4FGL J2035.4+1056   &   FSRQ   &   308.8431   &   10.9352   &   0.601\\
PKS 2052-47   &   4FGL J2056.2-4714   &   FSRQ   &   314.0682   &   -47.2466   &   1.489\\
PMN J2141-6411   &   4FGL J2141.7-6410   &   BCU   &   325.4435   &   -64.1874   &   0.959\\
OX 169   &   4FGL J2143.5+1743   &   FSRQ   &   325.8981   &   17.7302   &   0.211\\
PKS 2144+092   &   4FGL J2147.1+0931   &   FSRQ   &   326.7923   &   9.4963   &   1.113\\
PKS 2142-75   &   4FGL J2147.3-7536   &   FSRQ   &   326.803   &   -75.6037   &   1.139\\
PKS 2149-306   &   4FGL J2151.8-3027   &   FSRQ   &   327.9813   &   -30.4649   &   2.345\\
PKS 2155-304   &   4FGL J2158.8-3013   &   BL Lac   &   329.7169   &   -30.2256   &   0.1167\\
PKS 2155-83   &   4FGL J2201.5-8339   &   FSRQ   &   330.5802   &   -83.6366   &   1.865\\
4FGL J2202.7+4216   &   4FGL J2202.7+4216   &   BL Lac   &   330.6804   &   42.2778   &   0.0668\\
CTA 102   &   4FGL J2232.6+1143   &   FSRQ   &   338.1517   &   11.7308   &   1.032\\
PKS 2233-148   &   4FGL J2236.5-1433   &   BL Lac   &   339.142   &   -14.5562   &   0.325\\
TXS 2241+406   &   4FGL J2244.2+4057   &   FSRQ   &   341.053   &   40.9538   &   1.171\\
PMN J2250-2806   &   4FGL J2250.7-2806   &   BL Lac   &   342.6854   &   -28.1109   &   0.525\\
3C 454.3   &   4FGL J2253.9+1609   &   FSRQ   &   343.4906   &   16.1482   &   0.859\\
PKS 2255-282   &   4FGL J2258.1-2759   &   FSRQ   &   344.5248   &   -27.9726   &   0.926\\
PKS B2258-022   &   4FGL J2301.0-0158   &   FSRQ   &   345.2832   &   -1.9679   &   0.777\\
B2 2308+34   &   4FGL J2311.0+3425   &   FSRQ   &   347.7722   &   34.4197   &   1.817\\
PKS 2320-035   &   4FGL J2323.5-0317   &   FSRQ   &   350.8831   &   -3.2847   &   1.411\\
1ES 2322-409   &   4FGL J2324.7-4041   &   BL Lac   &   351.1861   &   -40.6804   &   0.174\\
PKS 2326-502   &   4FGL J2329.3-4955   &   FSRQ   &   352.337   &   -49.928   &   0.518\\
PMN J2345-1555   &   4FGL J2345.2-1555   &   FSRQ   &   356.3019   &   -15.9188   &   0.621\\
1ES 2344+514   &   4FGL J2347.0+5141   &   BL Lac   &   356.7702   &   51.705   &   0.044\\
PKS 2345-16   &   4FGL J2348.0-1630   &   FSRQ   &   357.0109   &   -16.52   &   0.576\\\hline
\multicolumn{6}{l}{Notes:}\\
\multicolumn{6}{l}{Column 1 is the name of the object;}\\
\multicolumn{6}{l}{Column 2 is the name of Fermi 4FGL;}\\
\multicolumn{6}{l}{Column 3 is the type of source (BL Lac or FSRQ);}\\
\multicolumn{6}{l}{Column 4 and Column 5 are right ascension (RA) and declination (Dec) coordinates;}\\
\multicolumn{6}{l}{Column 6 is the redshift of the source.}\\
\end{longtable}

\bsp	
\label{lastpage}
\end{document}